\documentclass{aastex}
\usepackage{float}
\usepackage{graphics}

\begin{document}


\title{Infrared Observations of Southern Classical Novae 1991 to 1992}

\author{Thomas E. Harrison, Joni J. Johnson}

\affil{Department of Astronomy, New Mexico State University, Box 30001, MSC 
4500, Las Cruces, NM 88003-8001}

\begin{abstract}
We report on a program to monitor classical novae (CNe) to determine if
they produced dust in the ejecta created by their outbursts. Of the 
ten systems we followed, five produced dust. We also present limited 
infrared and optical spectroscopy of these objects. We present a complete
$JHKLM$ spectrum for V992 Sco. V992 Sco was one of the brightest CNe in the
infrared of all time, and our $M$-band spectrum of this object shows strong
emission from the CO fundamental. We believe this to be the first, and
only, spectroscopic observation of this feature in a CNe.
\end{abstract}

\section{Introduction}

Classical novae (CNe) are thermonuclear runaways on a white dwarf. Mass is 
continually accreted by the white dwarf primary from a low mass companion. This 
material builds up in a shell on the white dwarf until the density and 
temperature reach a critical point, and a runaway is ignited (see Yaron et al. 
2005, and references therein). The resulting explosion drives a high velocity 
outflow, removing some of the accreted envelope of the white dwarf. In some 
subset of CNe, dust forms in the ejected shell.

There is a significant range in the outburst luminosities of classical novae 
(c.f., Shara et al. 2018), which appears to be related to the speed of the 
evolution of the 
thermonuclear runaway, which in turn has been tied to the mass of the white 
dwarf primary. The higher the mass of the white dwarf, the smaller the amount 
of material that needs to be accreted to ignite a thermonuclear runaway. This 
results in smaller ejecta masses, higher ejecta velocities, and a rapid burning 
and exhaustion of the (remaining) accreted shells (see Starrfield et al. 2008). 
Thus, the most luminous CNe have the shortest outbursts.

The maximum magnitude-rate of decline (MMRD, e.g., Della Valle \& Livio 1995) 
relation has been used for decades to estimate the outburst luminosities of 
CNe. The MMRD relates the absolute visual magnitude at maximum to the time it 
takes to decline by two ($t_{\rm 2}$), or three ($t_{\rm 3}$) magnitudes
from visual maximum. Using parallaxes, Harrison et al. (2013) found 
that the MMRD did not accurately reproduce the outburst luminosities of the
four CNe they studied. In fact, all four of those objects exceeded the 
Eddington luminosity at their visual maxima, with the two fastest CNe (V603 Aql, 
GK Per), exceeding it by greater amounts than the two slower CNe (DQ Her, 
RR Pic). Shore (2013) shows that the asphericity of the ejecta, combined
with the ejecta masses, filling factors and velocities all influence the
MMRD. Harrison et al. found that archival {\it micrometer} measurements
of the expansion of the CNe ejecta for these four CNe supported the conclusions 
of Shore.

While there are many unsolved issues in CNe research (see the review by Shara 
2014), the one that we find the most interesting is why some CNe produce dust, 
while others do not. For example, two of the CNe in the program discussed below, 
V838 Her (Nova Her 1991) and V868 Cen (Nova Cen 1991) both produced dust. V838 
Her was a very fast CNe, with $t_{\rm 3}$ = 5.3 d, while V868 Cen had 
$t_{\rm 3}$ $\sim$ 95 d. It is simple to detect the formation of dust in the 
ejecta of classical novae, as they suddenly develop an infrared excess. With 
monthly access to an infrared photometry system, we began a program to 
photometrically monitor southern classical novae that erupted during the years 
of 1991 and 1992. Here we present the results of that program.

\section{Observations}

\subsection{Near-Infrared Data}

Nearly all of the infrared data for our program were obtained using the 
Infrared Photometry System on the Siding Springs 2.3 meter telescope. The 
photometry was calibrated to a set standards that comprised the ``Mount Stromlo 
and Siding Spring Observatory system'' (McGregor 1994). The instrument used
was an InSb photometer on a telescope with a chopping secondary. For the $JHK$ 
photometry a 10" aperture with a 20" north-south throw was used. For some of 
the $L'$-, and for all of the $M$-band photometry, a 7" aperture was used. The 
telescope was repeatedly nodded back and forth between the two beam positions. 
Additional $JHK$ data was obtained on 1998 May 29 using 
CIRIM\footnote{http://www.ctio.noao.edu/instruments/ir\_instruments/cirim/cirim.html} on the CTIO 1.5 m telescope.
CIRIM was an infrared imager using a 256 square HgCdTe array.

The infrared photometry for all of the targets of our program is presented
in Table \ref{irphot}. Note that the data set presented here actually represents 
{\it 105 nights} of assigned time on the 2.3 m telescope at Siding Springs 
Observatory (SSO).  In those bad old days, we required photometric conditions to 
obtain data with this system, and for us, this occurred less than 20\% of the 
time. It is also true, however, that several of the targets we observed would 
have been {\it much too bright} to have been observed using a modern array-based 
system on a 2.3 m telescope. In addition, with a chopping system, we could 
observe the brighter objects during the day.

Infrared spectra were obtained using the Cooled Infrared Grating Spectrometer 
(CIGS; Jones et al. 1982) on the 2.3 m telescope. To provide for telluric 
correction, observations of $\alpha$ Cen were used. During 1991, CIGS had a 
single InSb detector, and was a scanning/chopping spectrograph. As in the old 
mode of photometry, the secondary mirror was chopped on and off the source at a 
frequency of 10 Hz, and the ``sky subtracted'' value at each grating position 
was stored. Like the photometry, we observed in a ABBA pattern, nodding the 
telescope back and forth after each complete wavelength scan of the grating. It 
took 45 minutes to obtain a single ABBA set of spectra in each bandpass, 
regardless of target brightness. It required nearly photometric conditions 
for two hours to achieve a spectrum covering all of $JHK$, something quite rare 
for our observing runs at SSO. Thus, even though several of the novae 
of 1991 were bright enough to have been observed with CIGS, we rarely had the 
necessary conditions for spectroscopic campaigns. V838 Her (Harrison \& 
Stringfellow 1994) was the rare exception. During the latter half of 1991, we
replaced the single element detector in CIGS with a linear array of 12 InSb 
detectors, resulting in much more efficient operation. 

\subsection{Optical Photometry and Spectroscopy}

In conjunction with the infrared program, we also obtained supporting optical 
photometry and spectroscopy. The photometric data was obtained on the SSO 1 m 
telescope using a variety of CCD cameras, all of them with the standard Bessell 
UBV(RI)$_{\rm c}$ filter set. These data are essential for producing spectral 
energy distributions (SEDs) for dust shell modeling. Optical spectra were 
obtained using the 74" telescope located at Mount Stromlo using the Cassegrain 
spectrograph. These data are useful to monitor the excitation state of the 
ejecta so as to infer the temperature of the central source, another quantity 
needed in modeling the dust shells of CNe. We also present spectra for V838 Her 
and V868 Cen obtained using 
the Dual-Beam Spectrograph (DBS\footnote{http://rsaa.anu.edu.au/observatories/instruments/dual-beam-spectrograph-dbs}) on the 2.3m telescope at Siding Springs.

\section{Results}

We discuss the individual novae below, ordered (mostly) by quantity of data
and scientific return, but also by year observed.

\subsection{V838 Her}

Our results for V838 Her (Nova Herculis 1991) have been presented in Harrison 
\& Stringfellow (1994). We were extremely lucky to be at the 2.3 m telescope 
when R. McNaught
called us and asked us if we were interested in a new nova that had just
been discovered. We actually used the 2.3 m telescope to refine the position
of the nova (IAUC 5222), and were able to obtain photometry within ``3 days
of outburst''  (1 day after discovery). What we did not present in our
paper on the infrared development of V838 Her was a magnificent spectrum
obtained for us by M. Gregg on 1991 April 13 using the DBS on the 2.3 m 
telescope. In Fig. 1 we present those data along with an inset 
showing the complex structure of the H$\alpha$ line, 21.7 d after the initiation 
of the outburst.

\renewcommand{\thefigure}{1}
\begin{figure}
\centerline{{\includegraphics[width=12cm]{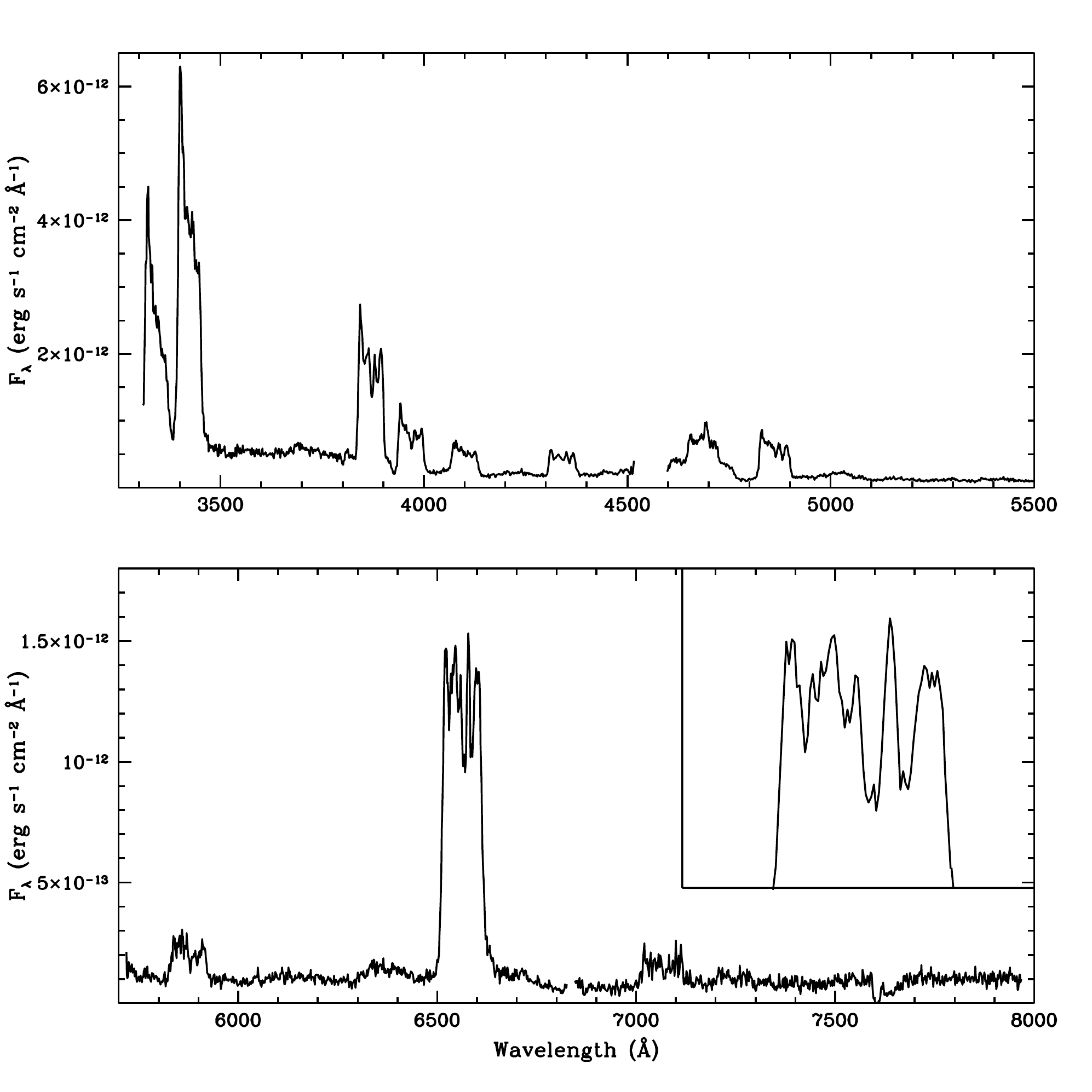}}}
\caption{The spectrum of V838 Her obtained on 1991 April 13 using the
DBS on the SSO 2.3 m telescope. The DBS places segments of spectra
on four different detectors, leading to small gaps in wavelength coverage.
Given this was the only DBS data set we ever reduced, the flux calibration
may be suspect. The inset presents a blow-up of the top of the H$\alpha$
line for examining its complex profile.}
\label{nher91}
\end{figure}

Because (at that time) the DBS used microchannel plates instead
of CCDs, it had excellent blue response. The DBS spectrum of V838 Her extends
all the way to 3300 \AA, and shows two very strong emission lines centered
at 3347 \AA, and 3423 \AA. Given that He I and O III were very weak at the
time of this spectrum, this suggests a low ionization state. There are a pair 
of Ne I lines at 3418 \AA ~and 3423 \AA, and a Ne II line at 3345 \AA. The 
strong line at 3871 \AA ~is probably [Ne III]. Besides this 
spectrum, some scattered optical photometry of V838 Her is listed in Table 
\ref{cneopt}.

\subsection{V868 Cen}

The most extensive data set we obtained for any of the CNe was for V868 
Cen (Nova Cen 1991). Bornak (2012) modeled these data using the Monte Carlo 
radiative transfer code ``DIRTY'' by Gordon et al. (2001). DIRTY allows for 
complex dust 
shell geometries to be constructed. Bornak investigated three geometries:
a homogeneous spherical shell, an inhomogeneous (clumpy) spherical shell, and
an equatorial torus. Bornak found that a dusty torus with large
grains ($a$ = 0.3 $\mu$m) provided the best match to the data. She found that 
the dust grains did not appear to evolve in size over the outburst. She derived
a total dust mass of 10$^{\rm -7}$ M$_{\sun}$. We have decided to partially 
redo this analysis, based upon the large data set that Bornak compiled. We 
present the optical data set she assembled from our observations made at SSO, 
as well as those collected from the literature, in Table \ref{cenphot}. 

Both Bornak and Williams et al. (1994) discuss the outburst of V868 Cen.
It is likely that the initial visual maximum was missed. The last
pre-discovery observation occurred some two weeks before discovery. The
nova was already declining when discovered. Bornak estimates that the original 
visual maximum reached $V$ $\sim$ 9.5 and occurred near JD 2448346 $\pm$ 2. We
will call this ``t$_{\rm 0}$'' (column 2 in Table \ref{cenphot}) in what
follows. The initial decline was rapid, with $t_{\rm 2}$ $\sim$ 6 d. 
However, it did not reach to three magnitudes until much later: 
$t_{\rm 3}$ $\sim$ 95 d. Using Downes \& Duerbeck (2000), we derive 
M$_{\rm V}$ = $-$7.0 with this $t_{\rm 3}$.
We present the evolution of the optical photometry in Fig. \ref{cenoptplot}.
As we will derive momentarily, V868 Cen is heavily reddened, and Bornak
found that visual estimates (e.g., those by the AAVSO) were offset to fainter 
magnitudes than the true $V$ value. The corresponding diagram for the
infrared data set is presented in Fig. \ref{cenirplot}.

\renewcommand{\thefigure}{2}
\begin{figure}
\centerline{{\includegraphics[width=12cm]{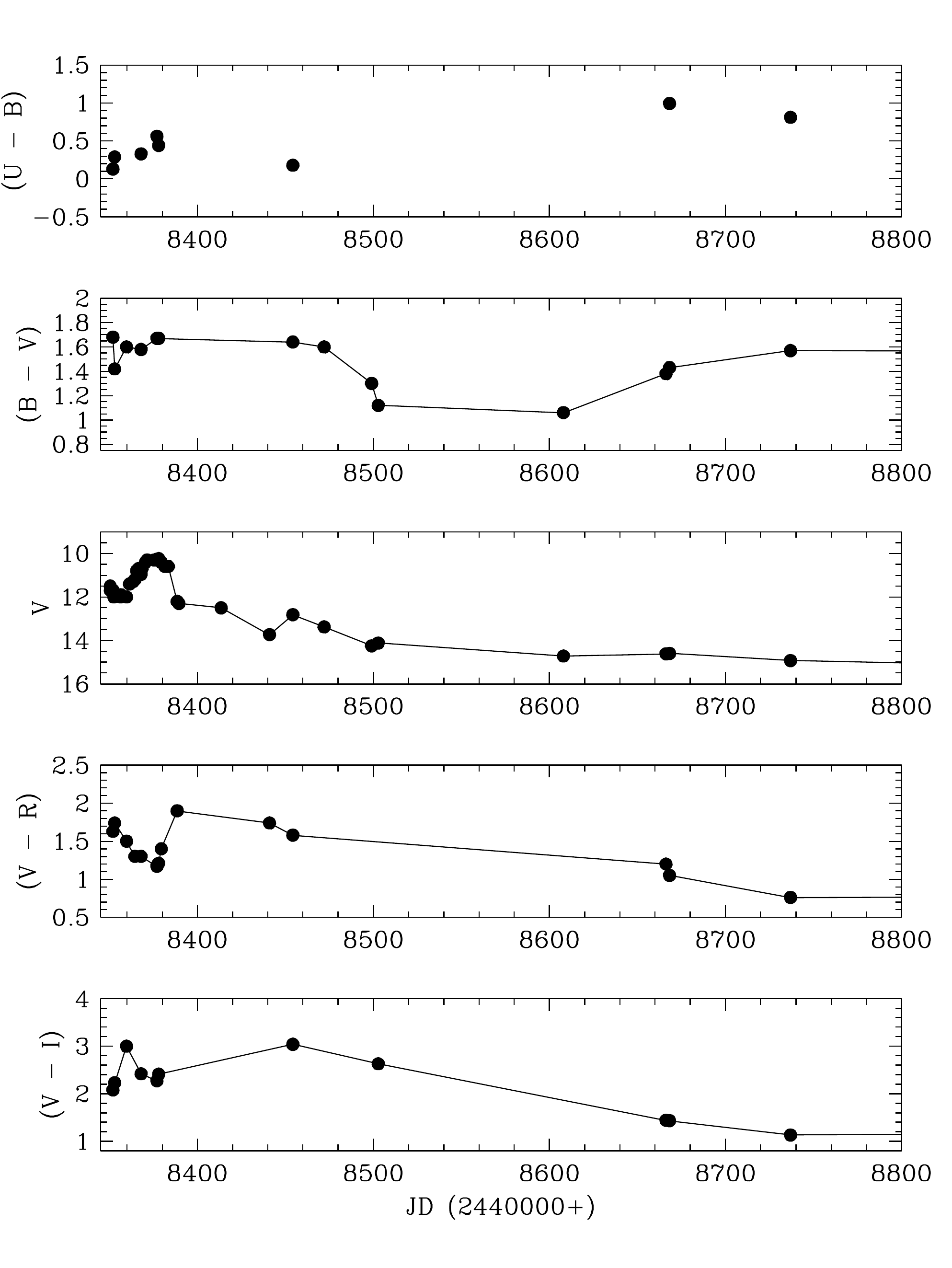}}}
\caption{The evolution of optical photometry of V868 Cen.}
\label{cenoptplot}
\end{figure}
\renewcommand{\thefigure}{3}
\begin{figure}
\centerline{{\includegraphics[width=12cm]{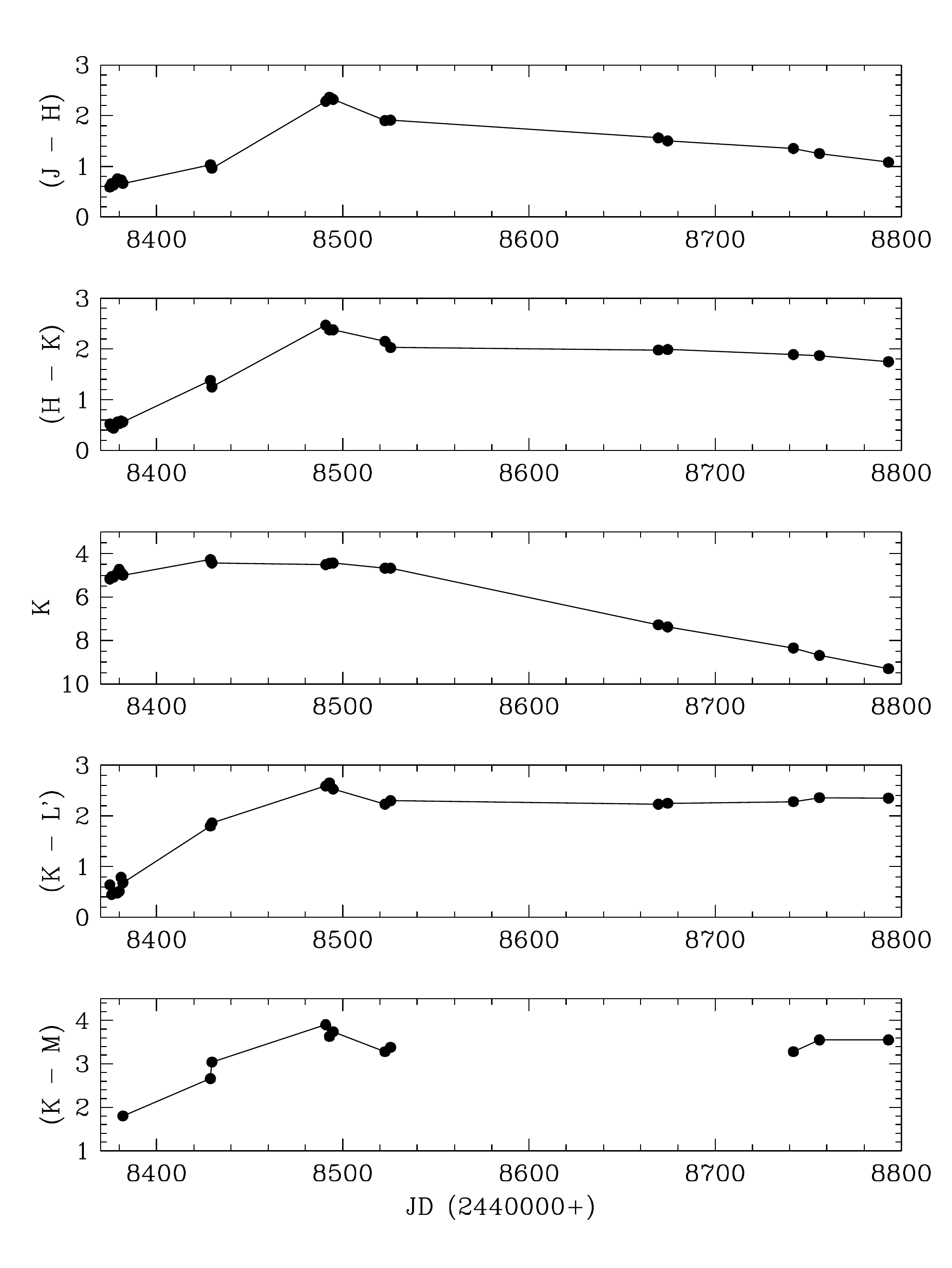}}}
\caption{The evolution of infrared photometry of V868 Cen.}
\label{cenirplot}
\end{figure}

\subsubsection{Extinction and Distance}

Bornak investigated the extinction in the region of V868 Cen by determining 
the color excesses for nearby stars with known spectral types and colors. Out 
to about 600 pc, the reddening is A$_{\rm V}$ $\sim$ 0.3 mag. Two open clusters 
located on a similar line-of-sight, at 1 kpc, have A$_{\rm V}$ $\sim$ 1.0 mag 
(NGC 5281 and NGC 5316). Additional evidence for a large extinction is given by 
van Genderen (1992), who published photometry for the B0 Ib supergiant
HR5171B located at a distance of 4.5 kpc, 39' NW of the CN, and it has 
A$_{\rm V}$ = 3.16.

We present the DBS spectrum of V868 Cen, Fig. \ref{cendbs}, obtained on
1991 April 12 (by M. Gregg).  These data were obtained three weeks after 
the secondary maximum. The spectrum is peculiar, with a blue hump centered 
near 4000 \AA, and a red continuum beyond 4500 \AA. Williams et al. (1994) 
classify V868 Cen as an ``Fe II'' nova, and emission lines due to Fe II are 
clearly present in the DBS spectrum. The blue end of the DBS spectrum has two 
strong lines that are not visible in the data presented by Williams et al. 
These are located at $\lambda$3935 \AA, and $\lambda$3970\AA. The NIST Atomic 
Spectra Data\footnote{https://physics.nist.gov/PhysRefData/ASD/lines\_form.html} 
has (strong) Fe I lines located at those positions. Bornak used the H I Balmer 
series to estimate a reddening using the ratios tabulated in Osterbrock (1989), 
and the Cardelli et al. (1989) reddening law. The average of the ratios of 
H I lines gave A$_{\rm V}$ = 3.4. Being that the DBS observation was obtained 
within a few weeks of a visual maximum, it is likely that the hydrogen lines 
were too optically thick for Case B conditions to apply.

\renewcommand{\thefigure}{4}
\begin{figure}[h]
\centerline{{\includegraphics[width=10cm]{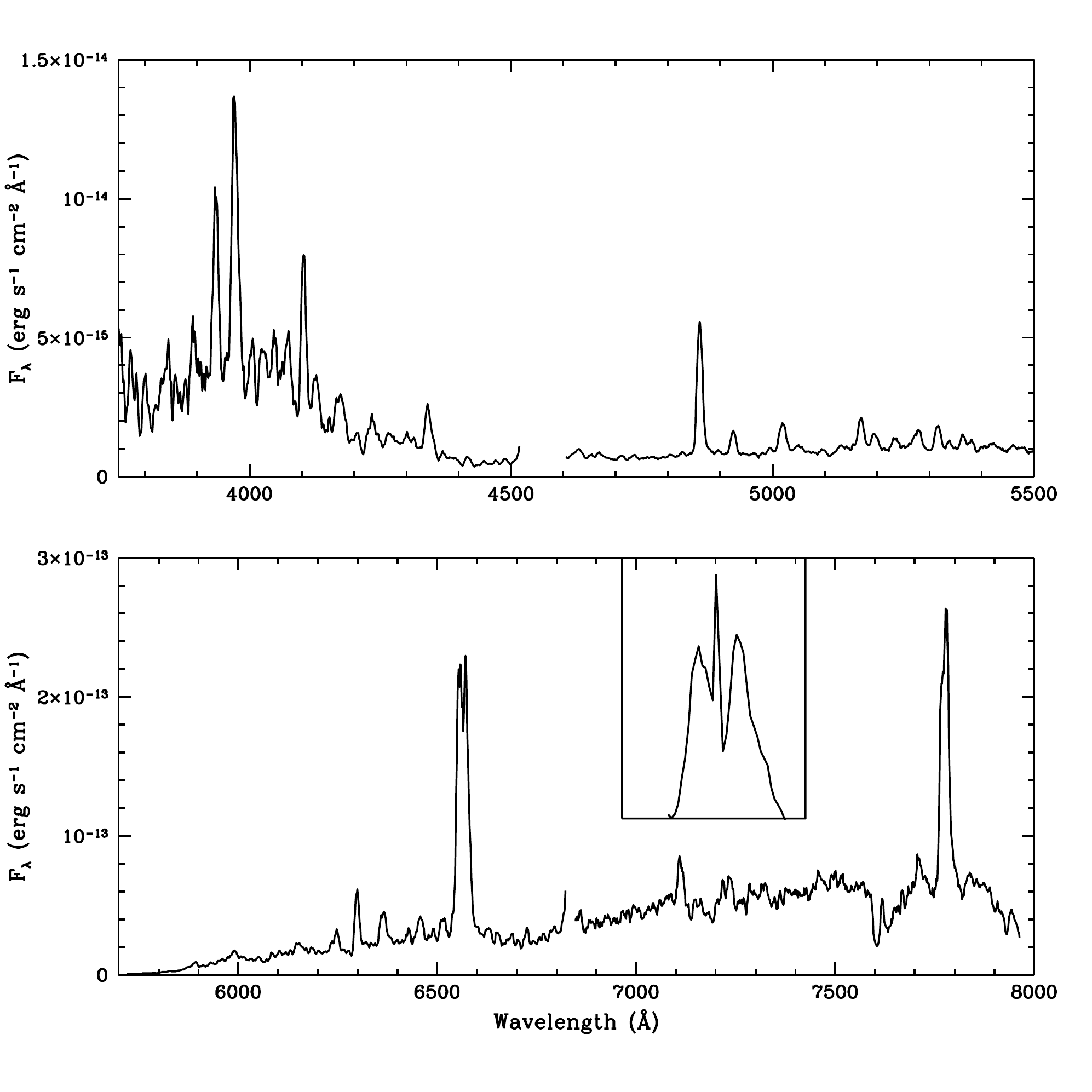}}}
\caption{The DBS spectrum of V868 Cen obtained on 1991 April 13 using the
DBS on the SSO 2.3 m telescope. The spectra have been boxcar smoothed
by five pixels, except for the inset of the H$\alpha$ line.}
\label{cendbs}
\end{figure}

\renewcommand{\thefigure}{5}
\begin{figure}[h]
\centerline{{\includegraphics[width=10cm]{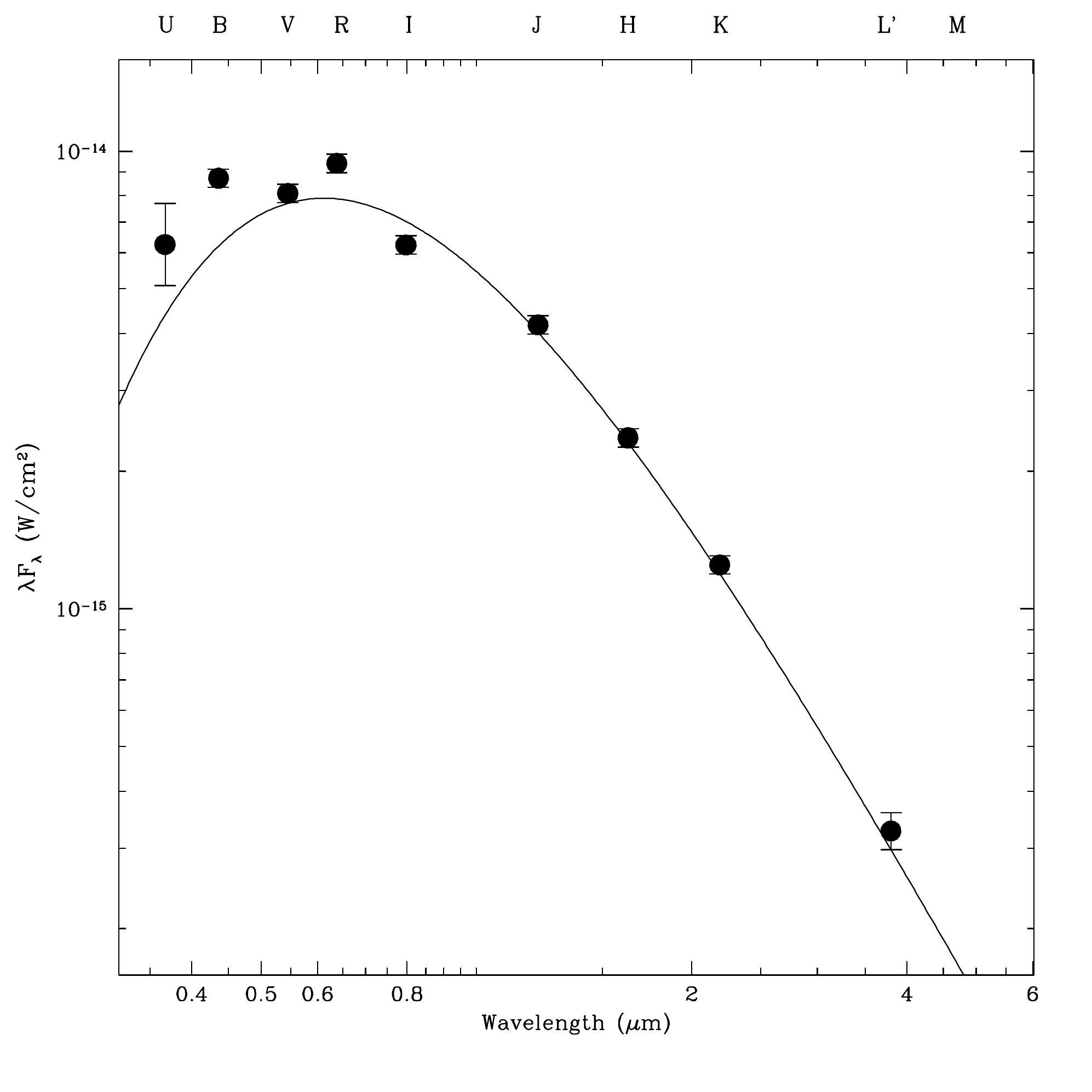}}}
\caption{The SED of V868 Cen for JD 2448377. The optical data is from
Table \ref{cenphot}, and near-IR data from Table \ref{irphot}. The error
bars are typical for the data in this paper, and will only appear in this
plot. The solid line is a black body with
T$_{\rm eff}$ = 6000 K reddened by A$_{\rm V}$ = 4.3 mag.}
\label{jd8377}
\end{figure}
Two other estimates use the ($B - V$) color. At the secondary maximum, we
have ($B - V$) = 1.65. Assuming the color of an F0 star, the color excess
is 1.35 mag, and A$_{\rm V}$ = 4.2 mag. Seven years after outburst, our CTIO 
observations showed that V868 Cen had ($B - V$) = 1.43. At this time we can
assume the underlying cataclysmic variable has settled to its quiescent state, 
and its visual light was dominated by emission from the accretion disk, which 
we presume has $(B - V$) $\sim$ 0. This gives A$_{\rm V}$ 
= 4.4 mag.  We find below that the average of these two, A$_{\rm V}$ = 4.3 mag,
 works well. If V868 Cen did reach $V$ = 9.5 at maximum, then its distance
is 2,750 pc.

\subsubsection{Modeling the Dust Shell of V868 Cen}

We begin our modeling on JD 2448377 (1991 April 30), near the peak of the 
secondary visual maximum, and at a time when we have near-simultaneous 
$UBVRIJHKL'$ photometry. These data, dereddened by A$_{\rm V}$ = 4.3 mag, are 
plotted in Fig. \ref{jd8377}. We have fit these data with a simple 
blackbody
spectrum with T$_{\rm eff}$ = 6000 K. There are four points that deviate from 
this blackbody fit. We can understand the deviation of the $U$, $B$, and $R$ 
photometry by referring to the DBS spectrum of V868 Cen. Obviously, the 
$R$-band is affected by the very strong H$\alpha$ emission line. The 
deviation of the $U$ and $B$ fluxes are due to the unusual blue ``hump'' that is 
present in the DBS spectrum. The origin of this hump is not obvious. None
of the spectra presented by Williams et al. (1994) for any of the CNe they
observed show such a feature. We presume it is due to iron. It was certainly 
not present in the spectroscopy we obtained seven weeks later.

Why the $I$-band data point appears to be slightly depressed is only hinted
at in the DBS spectrum of V868 Cen, but can be more clearly be seen in
the spectra presented by Williams et al. (1994). As can be seen in their Fig. 
5., the red continuum at this time is very complex, and appears to have broad 
absorption features. A down turn {\it is} visible right at the red end of our 
DBS spectrum. Thus, the SED for JD8377 is easily explained and simple to 
model.

The next set of near-IR data we want to model is that for JD 2448428. V868
Cen had declined by two magnitudes from its secondary maximum by this time.
The optical colors do not show much evolution, so we estimated their values
for modeling. The resulting SED is shown in Fig. \ref{jd8428}. The infrared
colors are peculiar with a very bright $M$-band detection, and a similar
``excess'' at $J$. We were able to use CIGS to obtain a full $K$-band spectrum
of the nova at this time, as well as partial $J$- and $H$-band spectra
(Fig. \ref{cigs}). The $H$- and $K$-band spectra reveal a very red continuum, 
while the $J$-band is flat with
an incredibly intense H I Pa$\beta$ emission line. The infrared excess,
and possible CO fundamental emission in the $M$-band (see V992 Sco, below),
suggests that dust is forming in the ejecta.

\renewcommand{\thefigure}{6}
\begin{figure}[h]
\centerline{{\includegraphics[width=10cm]{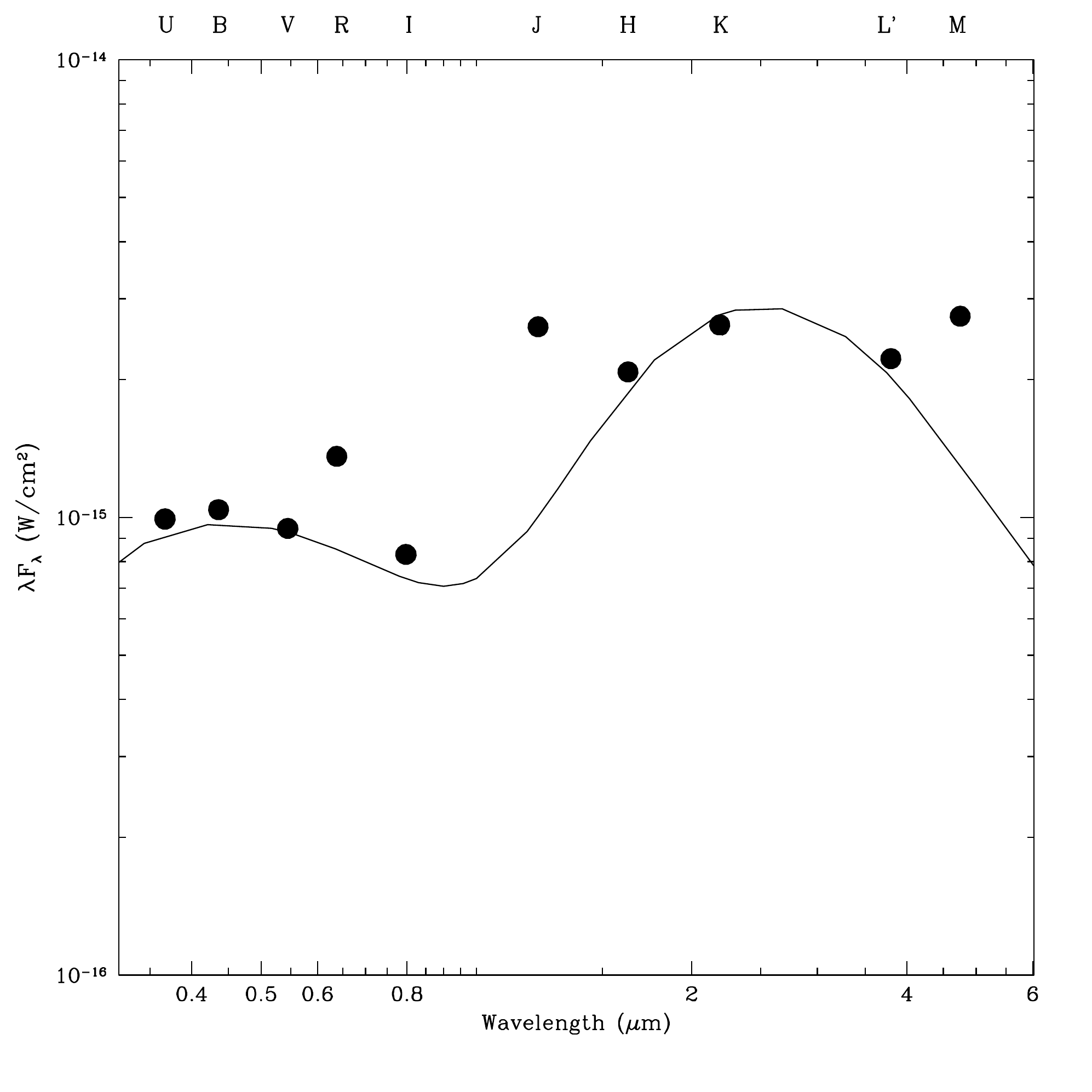}}}
\caption{The SED of V868 Cen for JD 2448428. The solid line is the
DUSTY model described in the text.}
\label{jd8428}
\end{figure}

\renewcommand{\thefigure}{7}
\begin{figure}[h]
\centerline{{\includegraphics[width=10cm]{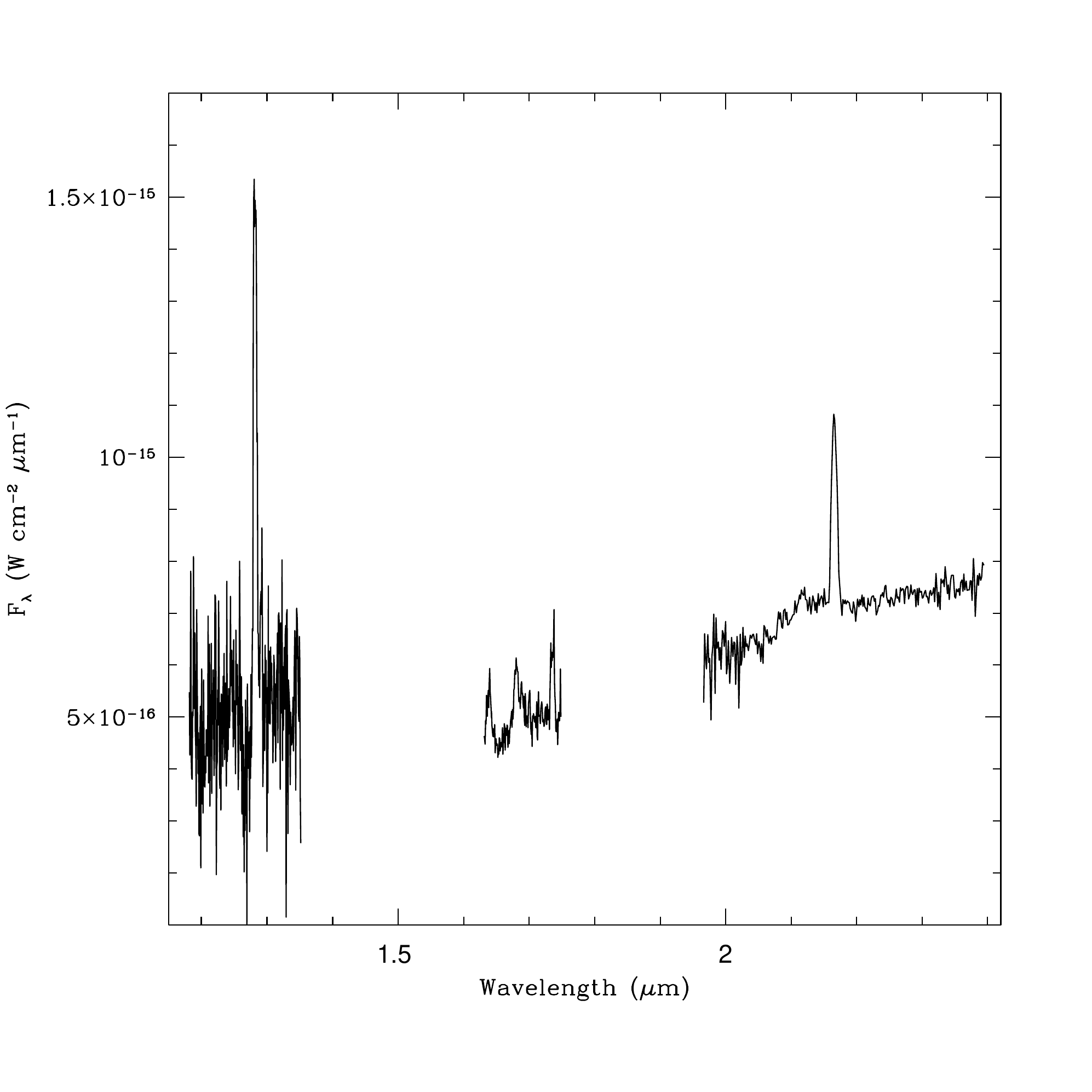}}}
\caption{The near-IR spectrum of V868 Cen on JD 2448428. The emission lines
in all three bandpasses are due to H I.}
\label{cigs}
\end{figure}

We used the dust shell radiative transfer code DUSTY (Iveni\'{c} et al. 2012) 
to model the SED of V868 Cen. To keep things simple, we used
the default ``MRN'' grain size distribution (Mathis et al. 1977), but found 
that we had to alter it to have a larger maximum grain size of $a$ = 0.45 
$\mu$m (versus the standard of 0.25 $\mu$m) to fit the data. We used only 
amorphous carbon grains. 
We set the grain sublimation radius temperature to 1500 K, and the central
source blackbody temperature to 12250 K. Our model has the dust density falling
off as $r^{\rm -2}$ in a thin shell. The final model is the solid line in
Fig. \ref{jd8428}. The fit is reasonable, though it is difficult to understand
the large deviation of the $J$-band at this time. Perhaps there is also
very strong He I 10830 \AA ~emission along with strong H I emission that
is causing its flux to be three times that of the dust shell model. It is
clear, however, that dust began to form in the ejecta not too long after the 
sudden drop in the $V$ magnitude near JD 2448400.  

\renewcommand{\thefigure}{8}
\begin{figure}[h]
\centerline{{\includegraphics[width=10cm]{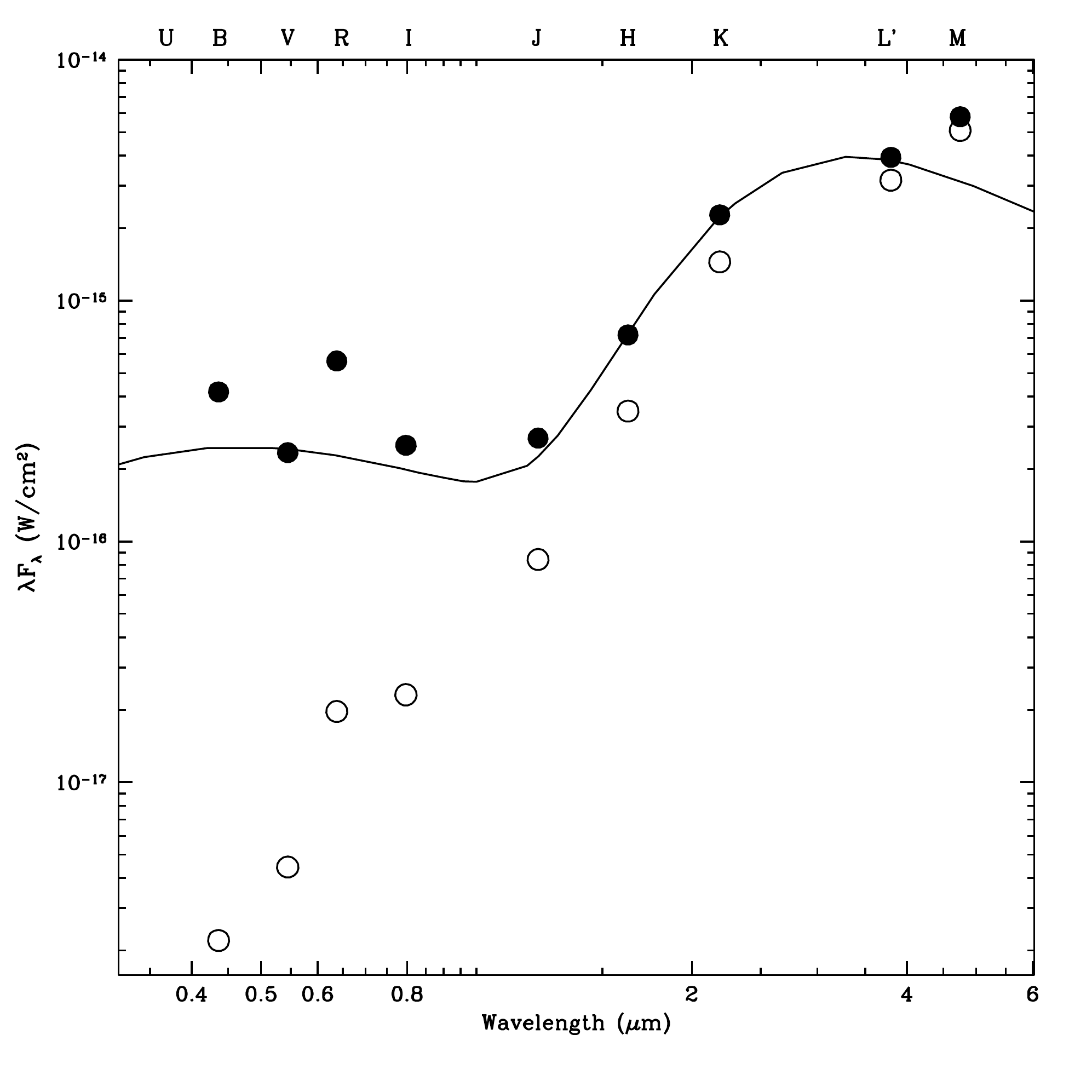}}}
\caption{The SED of V868 Cen for JD 2448500. The solid line is the
DUSTY model described in the text. The open circles are the unreddened data.}
\label{jd8500}
\end{figure}

For JD 2448500 we have near-simultaneous $UBVRIJHKL'M$ photometry. The
infrared SED is simpler now, Fig. \ref{jd8500}, except for the $M$-band excess. 
We use a nearly identical model as before (same MRN grain size distribution), 
but with a sublimation temperature for the grains of 1150 K. The thin shell
now has uniform density, and extends to twice the inner shell radius. The total
optical depth in the dust shell at this time is $\tau$ = 1.13. So, $\sim$ 1 mag
of visual decline since the previous observation is almost certainly due to the
thickening of the dust shell.

\renewcommand{\thefigure}{9}
\begin{figure}[h]
\centerline{{\includegraphics[width=10cm]{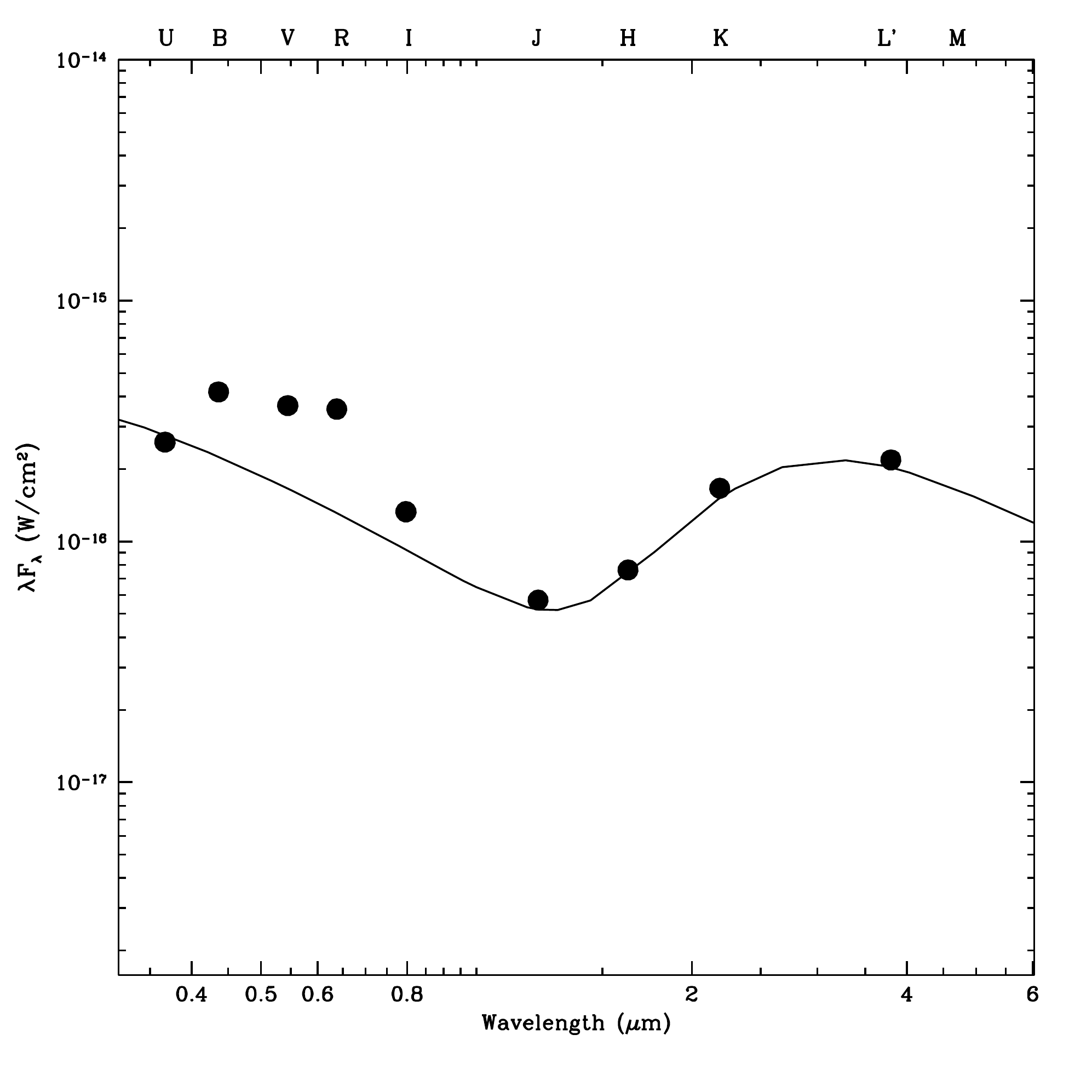}}}
\caption{The SED of V868 Cen for JD 2448674. The solid line is the
DUSTY model described in the text.}
\label{jd8674}
\end{figure}

The final data set we model is for JD 2448674, Fig. \ref{jd8674}, ten months
since outburst. The infrared SED is smooth, but the optical SED is complex. As 
shown in Fig. \ref{apr3092}, the optical spectrum of V868 Cen was dominated by 
line emission at these late stages. The main changes to the setup for DUSTY were
the central blackbody source temperature had to be much higher, 27500 K,
and the dust shell had to be much more extended, 2.3 times its inner
radius. Its density also had to fall of with radius, $\rho$ $\propto$ 
r$^{\rm -2}$.

\renewcommand{\thefigure}{10}
\begin{figure}[h]
\centerline{{\includegraphics[height=6cm]{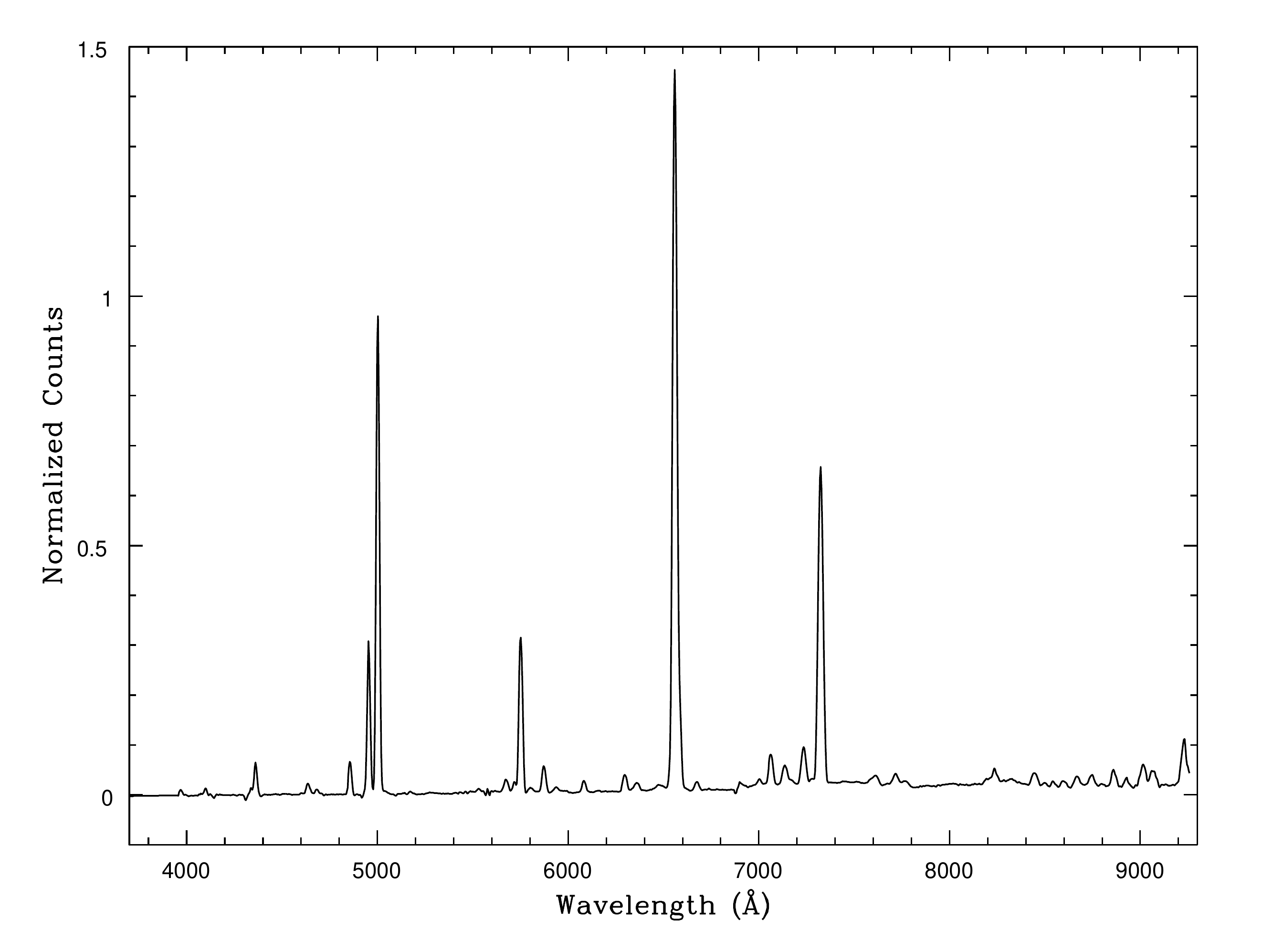}}}
\caption{The optical spectrum of V868 Cen for 1992 April 30 (JD 2448742).}
\label{apr3092}
\end{figure}

In conclusion, V868 Cen produced dust, but not very much dust, and at no
time did it deeply obscure the central source as seen in other
CNe. The low dust mass found by Bornak is certainly consistent with the
results found here, though our modeling is unable to actually constrain
the dust mass. Proper dust shell modeling requires optical photometry and 
spectroscopy with good temporal coverage to go along with the infrared data set
so that the origin of any deviant fluxes can be understood.

\subsection{V2264 Oph}

Williams et al. (1994) describe the discovery and decline rate of V2264 Oph
(Nova Oph 1991a), finding an approximate value of $t_{\rm 3}$ = 36 d. They 
classify it as an Fe II nova like V868 Cen, and thus maybe it too would produce 
dust. In Fig. \ref{v2264oph}, we present a subset of the infrared data showing 
the complex evolution of the near-IR SED of V2264 Oph that clearly shows the 
formation of dust 
in its ejecta. At the earliest times the SED resembled free-free emission.
By JD 2448380, however, an excess was developing in the $K$ and $L'$ bands. 
Just two weeks later, JD 2448398, there was a strong infrared excess in both $K$ 
and $L'$. When V2264 Oph was again accessible, JD 2448491, the SED had undergone 
an usual change: the ($K - L'$) color had gotten bluer by nearly one magnitude.

\renewcommand{\thefigure}{11}
\begin{figure}[h]
\centerline{{\includegraphics[height=10cm]{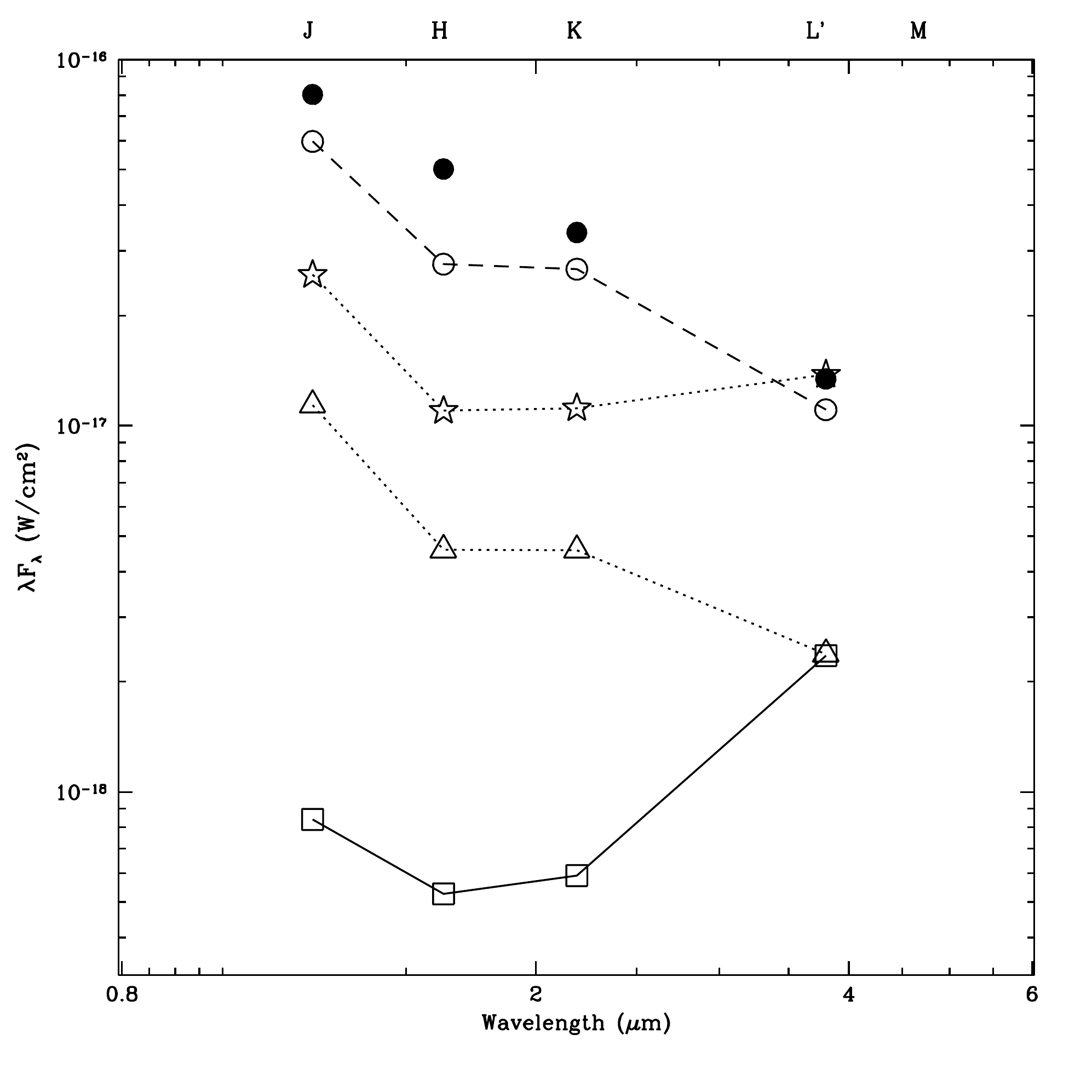}}}
\caption{The evolution of the SED for V2264 Oph. The solid circles are
for JD 2448375.10, the open circles are for 2448380.10, the stars are for
JD 2448397.99, the triangles are for JD 2448490.92, and the squares are for 
JD 2448796.15.}
\label{v2264oph}
\end{figure}

The final set of observations plotted in Fig. \ref{v2264oph}, for JD 2448796,
demonstrates the very slow fading of the $L'$ flux, in contrast with the other 
bandpasses (though
clearly an excess is present at $K$). Given that we were using an aperture
photometer, we examine whether source confusion might explain our $L'$-band
measurements: There are {\it no} sources in the $WISE$ $W1$ images within 
several arcminutes of V2264 Oph anywhere near as bright as what we observed for 
this source. Thus, the strong excess, and its unusual evolution, are real.
Unfortunately, V2264 Oph was never bright enough for us to detect it in the 
$M$-band. This, and the sparse optical data set, make it difficult to construct 
a dust shell model.

\subsection{V351 Pup}

The final CNe of 1991 for which we have reasonable photometric coverage is V351 
Pup (Nova Pup 1991). As described by Williams et al. (1993), this object was 
discovered by P. Camilleri on 1991 December 27, having $m_{v}$ = 6.4. Williams 
et al.  believe this must be close to the time of visual maximum. Our first 
infrared observations did not occur until 52 days later. Williams et al. classify
this object as another Fe II nova. The evolution of its SED over the four
months we observed it is presented in Fig. \ref{v351pup}. The SED is
consistent with being dominated by emission lines, as demonstrated by
the CIGS spectrum, Fig. \ref{v351cigs}, obtained on 1992 April 19 (JD 2448732). 
At no time is there any evidence for the production of dust.

\renewcommand{\thefigure}{12}
\begin{figure}[h]
\centerline{{\includegraphics[height=10cm]{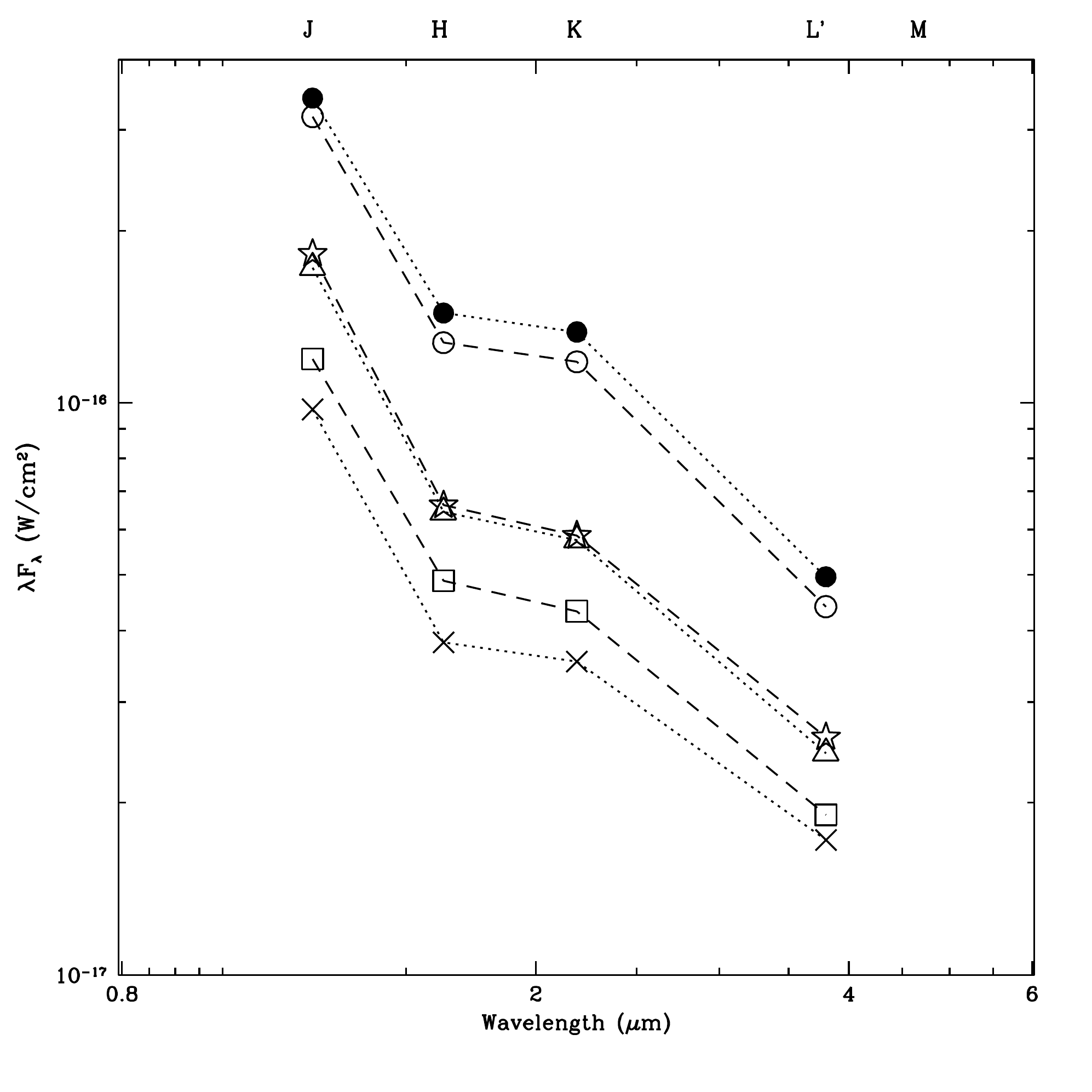}}}
\caption{The evolution of the SED for V351 Pup. All of the data in
Table 1 for this object are plotted except that for the penultimate date. }
\label{v351pup}
\end{figure}
\renewcommand{\thefigure}{13}
\begin{figure}[h]
\centerline{{\includegraphics[height=10cm]{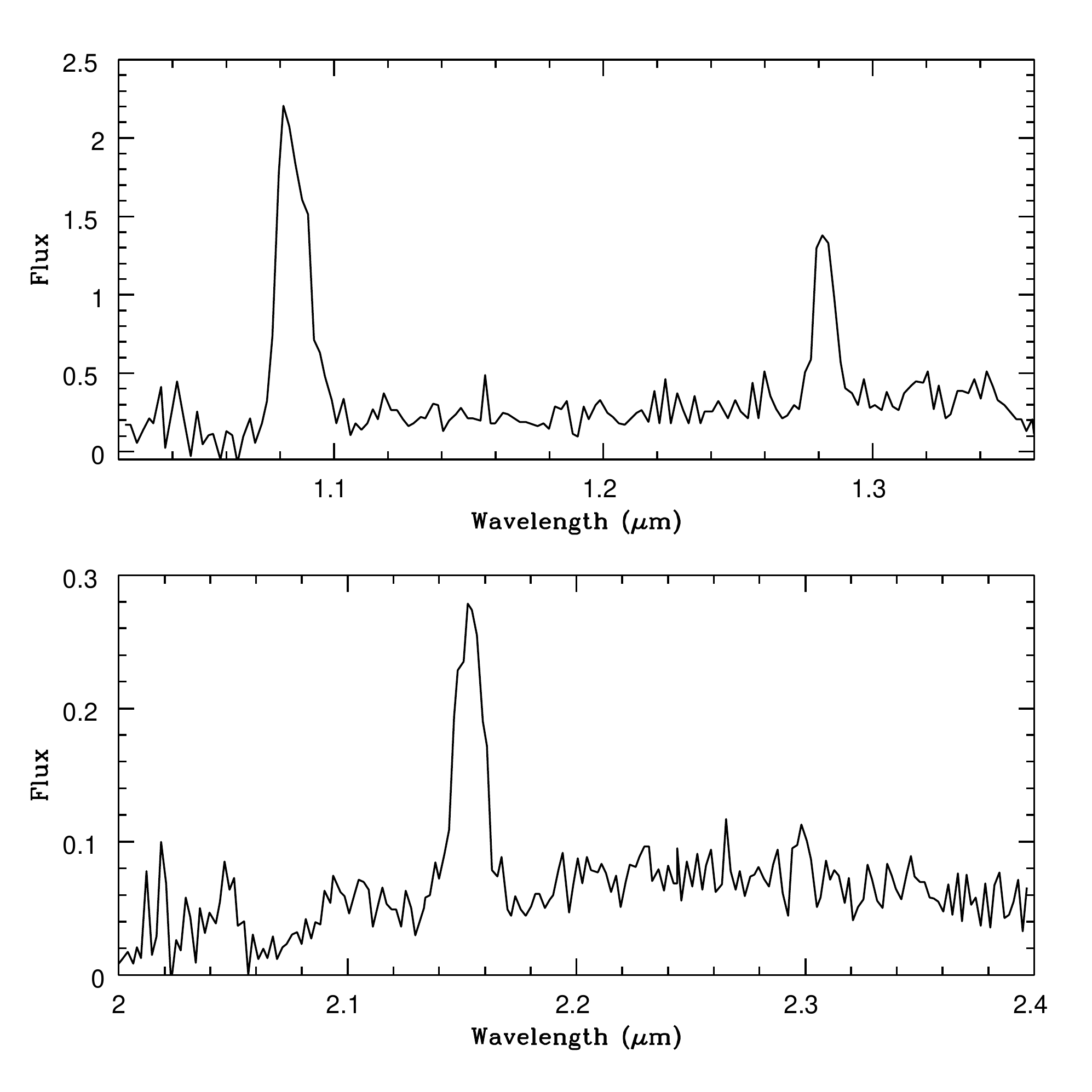}}}
\caption{The $J$- and $K$-band spectra of V351 Pup for JD 2448732. The raw
spectra have been divided by $\alpha$ Centauri, but have not been flux
calibrated.  }
\label{v351cigs}
\end{figure}

\subsection{The Other CNe of 1991: Nova Oph 1991b, V4160 Sgr, V444 Sct,
Nova LMC 1991a}

We only obtained a limited amount of data on these four CNe. We plot all of
the $JHKL$ data for Nova Oph 1991b, V4160 Sgr, and V444 Sct in Fig. \ref{n1991}.
Williams et al. (1993) classify V4160 Sgr as an He/N nova, with a rapid
decline: $t_{\rm 3}$ = 4 d. Our single epoch of observations show a 
continuum dominated by line emission. Our observations for V444 Sct spanned
four months, and hinted at an infrared excess at late times, however,
the $L'$ observation was just barely a 3$\sigma$ detection, and there is
no sign of an excess at $K$. V444 Sct was classified by Williams et al. as
an Fe II nova, so dust production would not be surprising, given the results
from above. Nova LMC 1991a was also classified as an Fe II nova. Its
SED (not plotted) is consistent with a continuum dominated by line emission.

\renewcommand{\thefigure}{14}
\begin{figure}
\centerline{{\includegraphics[height=8cm]{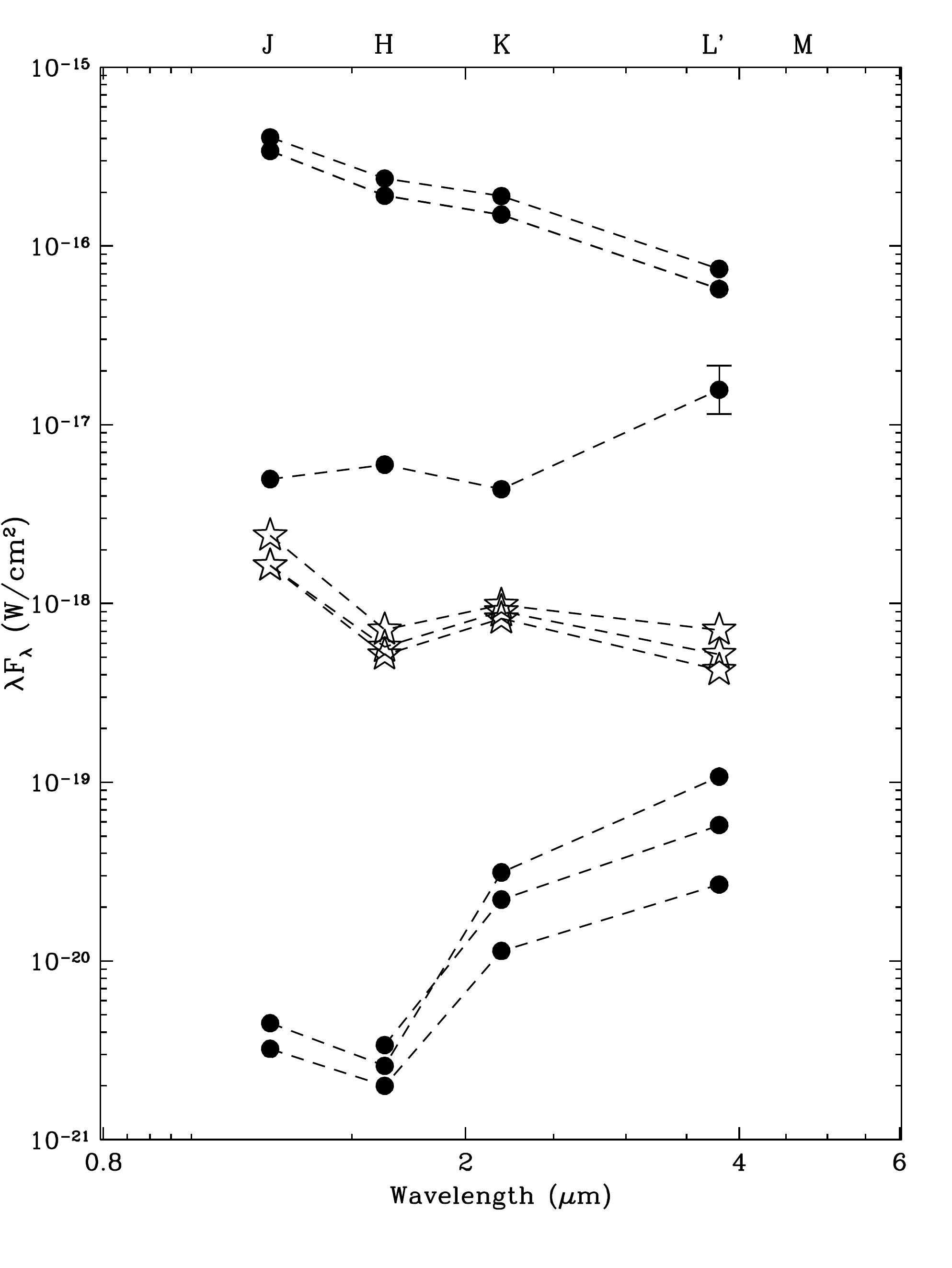}}}
\caption{The infrared SEDs of V445 Sct (filled circles, top), V4160 Sgr
(stars), and Nova Oph 1991b (filled circles, bottom). The fluxes for
V445 Sct and Nova Oph 1991b have been offset vertically for clarity.}
\label{n1991}
\end{figure}

Of these four objects, it is Nova Oph 1991b that is clearly the most 
interesting. The nova was found on a photograph taken on 1991 April 11 by P. 
Camilleri, but not discovered until 1991 October 28. At the time of the
photo it had a $m_{pv}$ = 9.3 (IAUC 5381). R. McNaught obtained $BVRI$ 
photometry and found that the object was still quite blue in early 
November of 1991. Our first observation did not occur until 1992 April 19.7. 
At that time it is clear that this object was enshrouded in a dust shell.
On 1992 April 29.6 we obtained an optical spectrum of this source (probably
the only one in existence), Fig. \ref{nophb}, showing an extremely strong 
[OIII] doublet. This doublet should be compared to the weak H$\beta$ line, 
demonstrating the large amount of extinction at this time. It is probably
time for Nova Oph 1991b to receive a proper variable star name.

\renewcommand{\thefigure}{15}
\begin{figure}
\centerline{{\includegraphics[height=8cm]{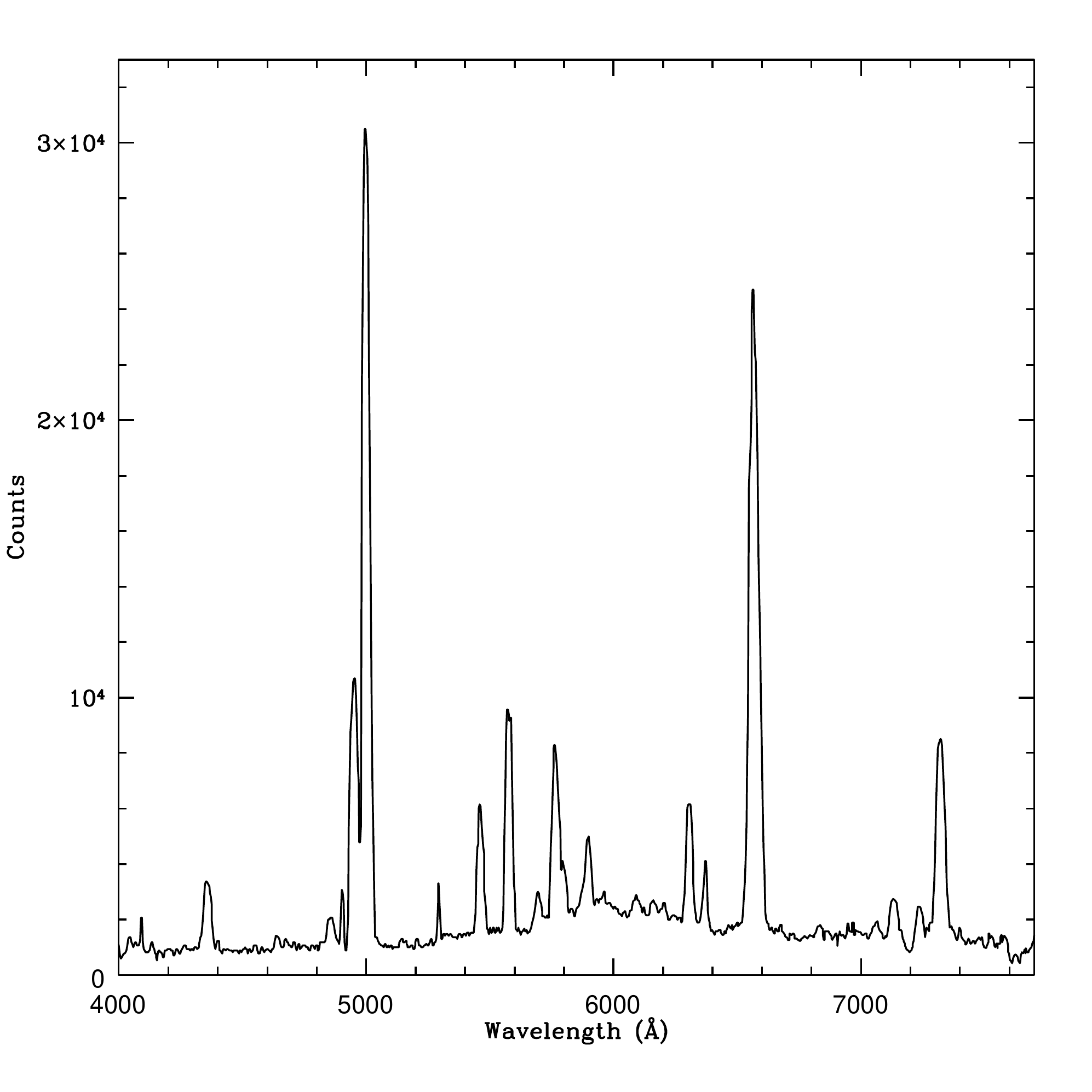}}}
\caption{The optical spectrum of Nova Oph 1991b obtained using the
Cassegrain spectrograph on the Mt. Stromlo 74" telescope on JD 2448742.10.}
\label{nophb}
\end{figure}

\subsection{The CNe of 1992: V992 Sco and V4157 Sgr}

\begin{flushleft}
{\bf V992 Sco:} Smith et al. (1995) discuss the complex light curve of V992 
Sco. Depending
on which measure you use, it is either a fast, or a slow nova. Using $t_{\rm 3}$
it is clearly a slow nova. V992 Sco was discovered by P. Camilleri on
1992 May 24; our first infrared observations were obtained one month later.
The $V$-band light curve (data in Table \ref{scophot}), and the evolution of the 
infrared photometry, are presented in Fig. \ref{v992irdata}. From the time of 
our first observations, the object got progressively redder, and redder. In the 
visual, the system fluctuated in brightness until JD 2448870, when it began 
what appears to have been a monotonic decline. 
\end{flushleft}

\renewcommand{\thefigure}{16}
\begin{figure}
\centerline{{\includegraphics[width=12cm]{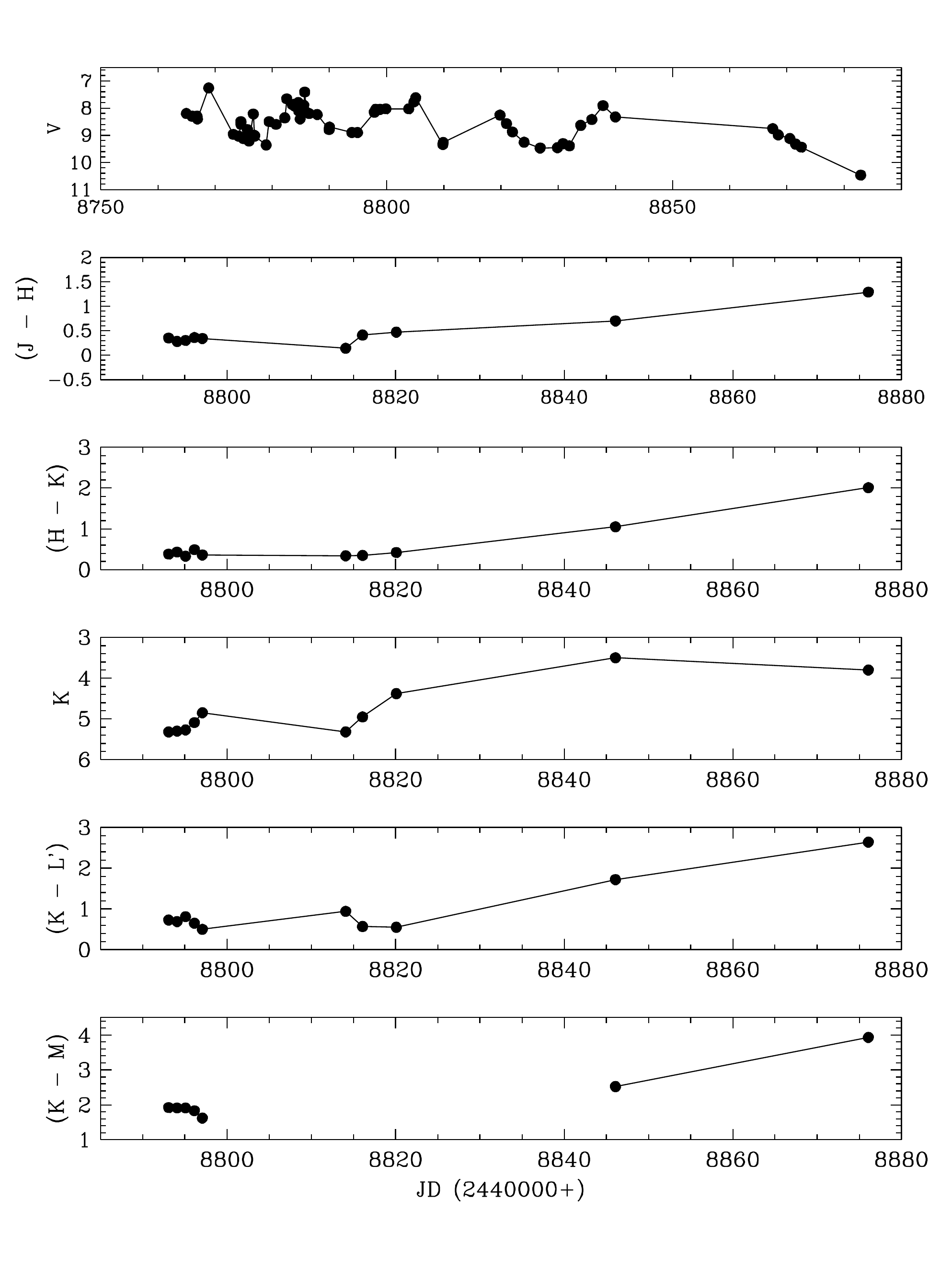}}}
\caption{The $V$-band light curve, and infrared photometry of V992 Sco.}
\label{v992irdata}
\end{figure}

Combining optical and infrared data, we model the SED of V992 Sco on
JD 2448795 in Fig. \ref{jd8795}. The $V$ through $K$-band SED is well-modeled
by a 6500 K blackbody (reddened by A$_{\rm V}$ = 1.4 mag), except for strong
excesses at $L'$ and $M$. In Fig. \ref{juneopt}, we present
an optical spectrum of V992 Sco obtained on JD 2448794. The spectrum at
this time is one of low excitation, with strong Ca II triplet emission,
and what appear to be weak P-Cygni profiles on some of the lines.

\renewcommand{\thefigure}{17}
\begin{figure}
\centerline{{\includegraphics[width=8cm]{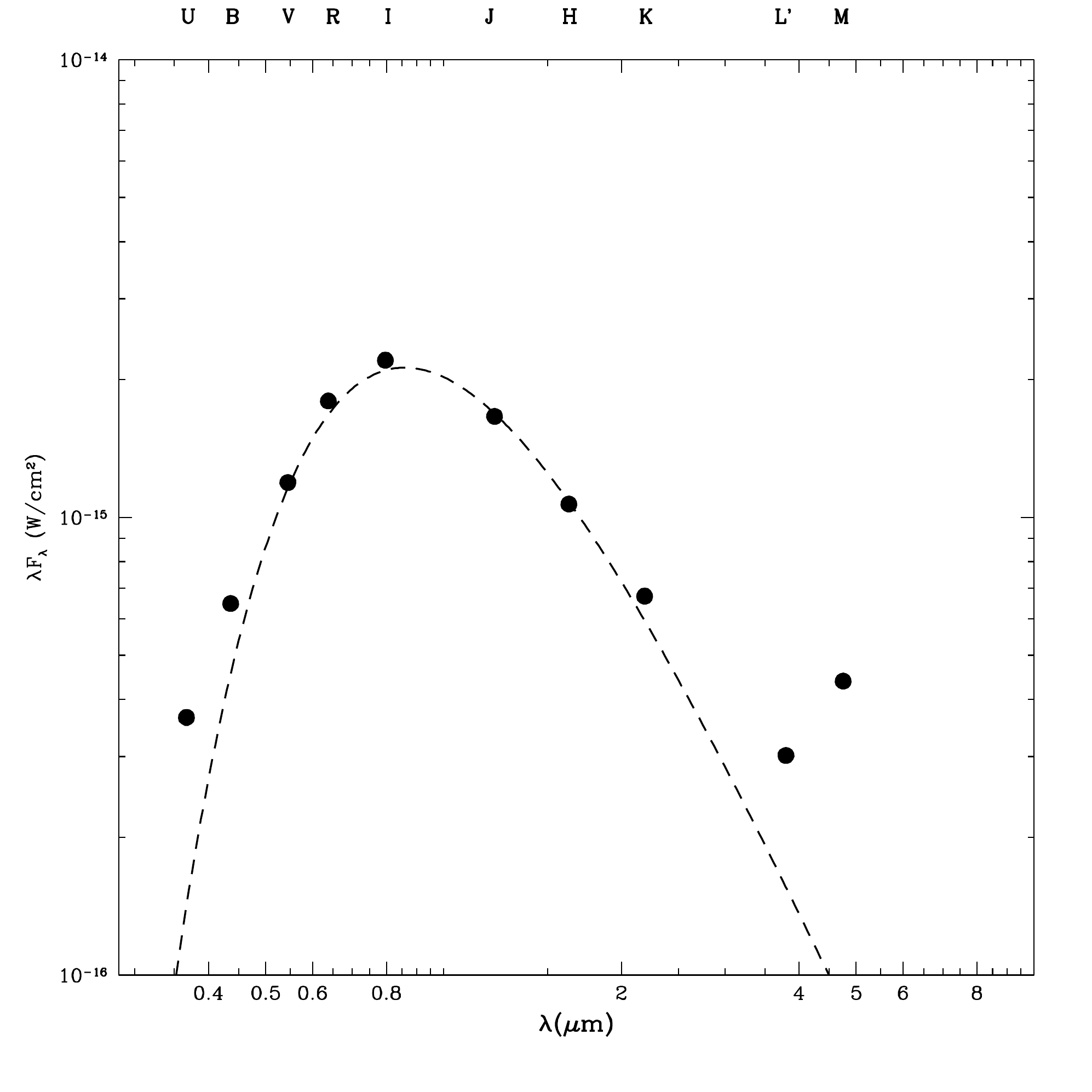}}}
\caption{The SED of V992 Sco on JD 2448795. The dashed line is a 6500 K 
blackbody reddened by A$_{\rm V}$ = 1.4 mag.}
\label{jd8795}
\end{figure}

\renewcommand{\thefigure}{18}
\begin{figure}
\centerline{{\includegraphics[width=8cm]{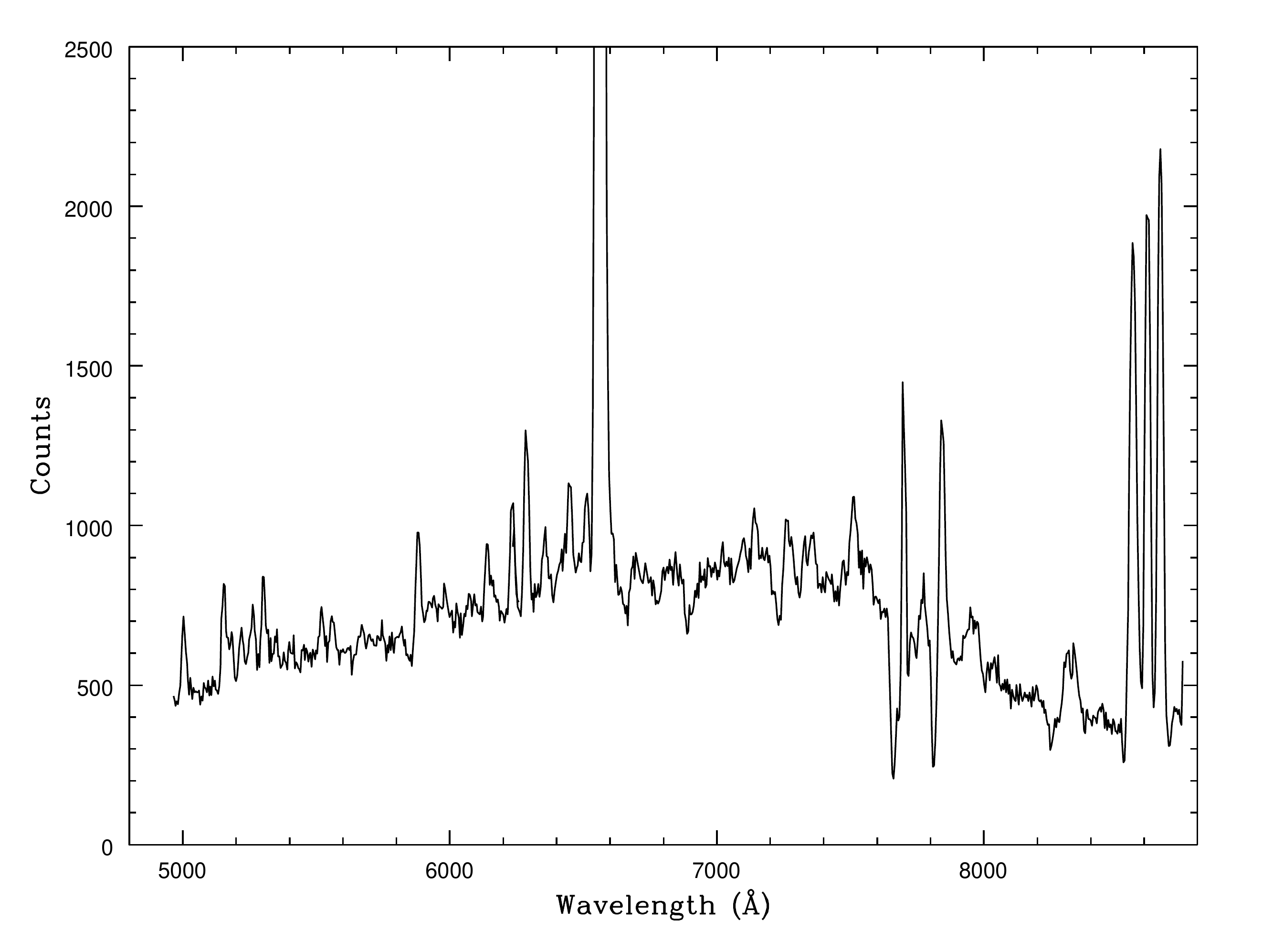}}}
\caption{The optical spectrum V992 Sco on JD 2448794, twenty four hours
earlier than the data presented in Fig. \ref{jd8795}.}
\label{juneopt}
\end{figure}

The origin of the $L'$-band excess can be seen in the spectrum obtained on 1992 
June 23, shown in Fig. \ref{jhklspec}: H I Br$\alpha$ emission at 4.05 $\mu$m. 
While the near-IR spectra are $mostly$ dominated by hydrogen lines, it is 
important
to note that in the $H$-band, H I Br10 (1.74 $\mu$m), and H I Br11 (1.68 $\mu$m),
are convolved with emission lines from C I. The second strongest line in
the $J$-band, located at 1.13 $\mu$m, is probably the O I/N I blend seen
in some CNe (e.g., Venturini et al. 2004). In the $K$-band we have the Na I 
doublet at 2.2 $\mu$m in emission. There is also an emission feature at
2.088 $\mu$m that we are unfamiliar with. On a scaled and smoothed spectrum,
it appears that the first overtone features of CO are in emission.

\renewcommand{\thefigure}{19}
\begin{figure}
\centerline{{\includegraphics[width=12cm]{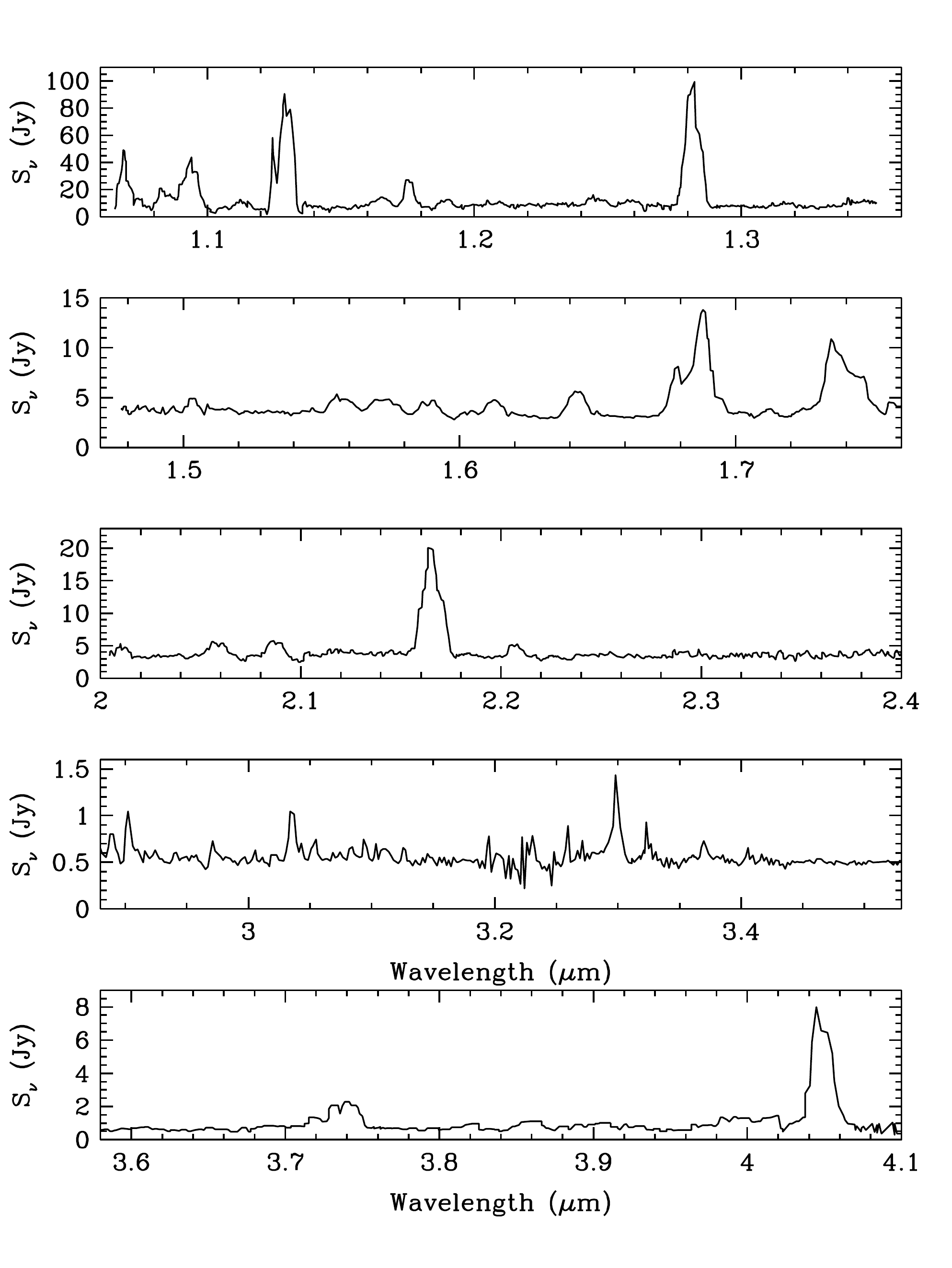}}}
\caption{The $JHKL$ spectrum of V992 Sco on 23 June 1992.}
\label{jhklspec}
\end{figure}

In the $L$-band, there is an emission line from C I at 3.37 $\mu$m, and
there may be a series of emission features from N I located near 3.30 $\mu $m, 
convolved with H I Pf$\delta$. The $M$-band spectrum is presented in Fig. 
\ref{mband}. This spectrum shows the CO fundamental in emission, the source of 
the excess seen in the SED for JD 2448795. The $M$-band excesses seen in many 
CNe are harbingers for the production of dust, and we will see that this is true
for V992 Sco.

\renewcommand{\thefigure}{20}
\begin{figure}
\centerline{{\includegraphics[width=7cm]{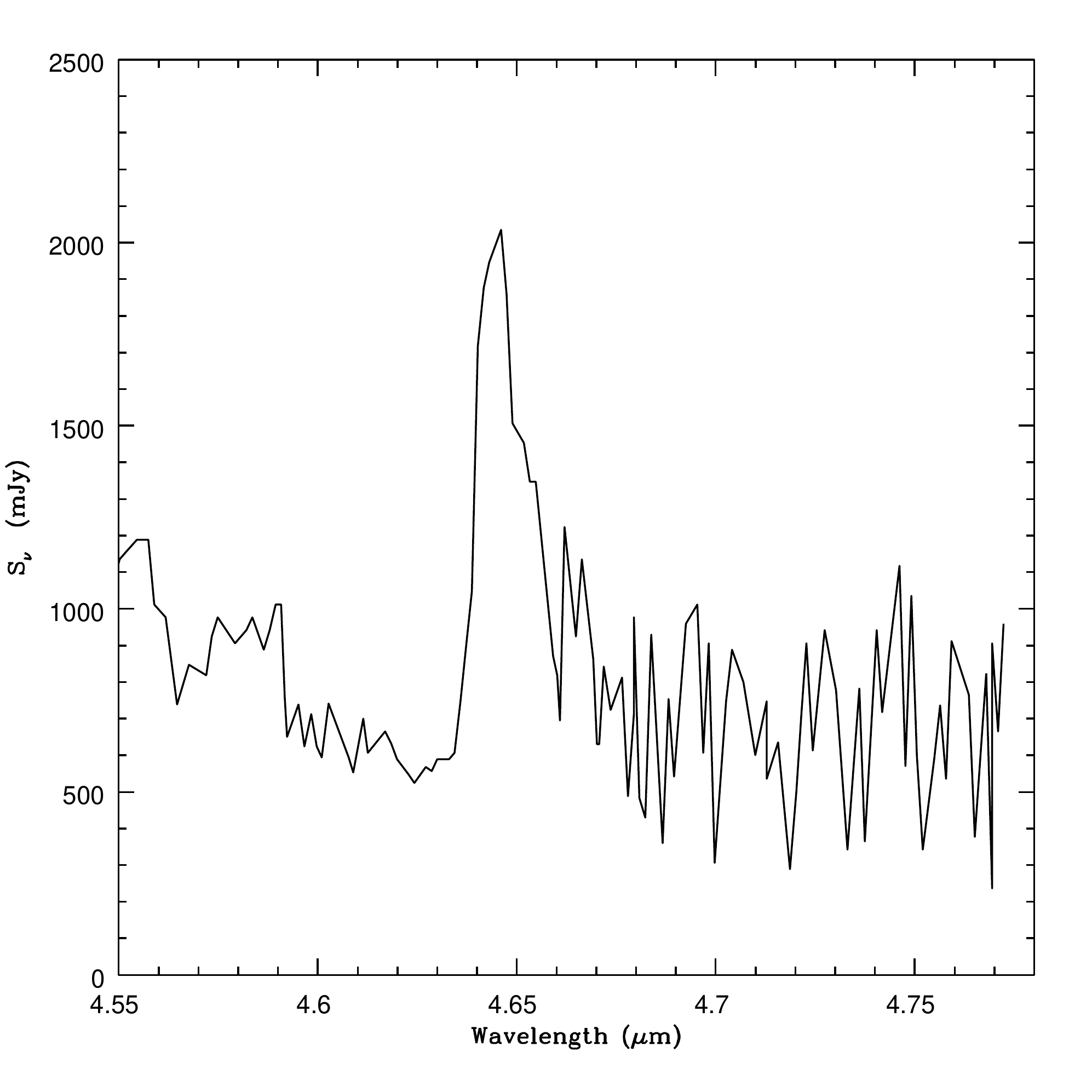}}}
\caption{The $M$-band spectrum of V992 Sco on 23 June 1992.}
\label{mband}
\end{figure}
In 1992 August we used the 74" telescope on Mt Stromlo to obtain some
moderate resolution spectroscopy of V992 Sco, Fig. \ref{scoaug}. All of the 
stronger emission lines, except those for the Ca II triplet, had developed deep 
P Cygni profiles. The ionization state of the ejecta remained low, though the 
He I lines were now stronger than seen in June.

\renewcommand{\thefigure}{21}
\begin{figure}
\centerline{{\includegraphics[width=8cm]{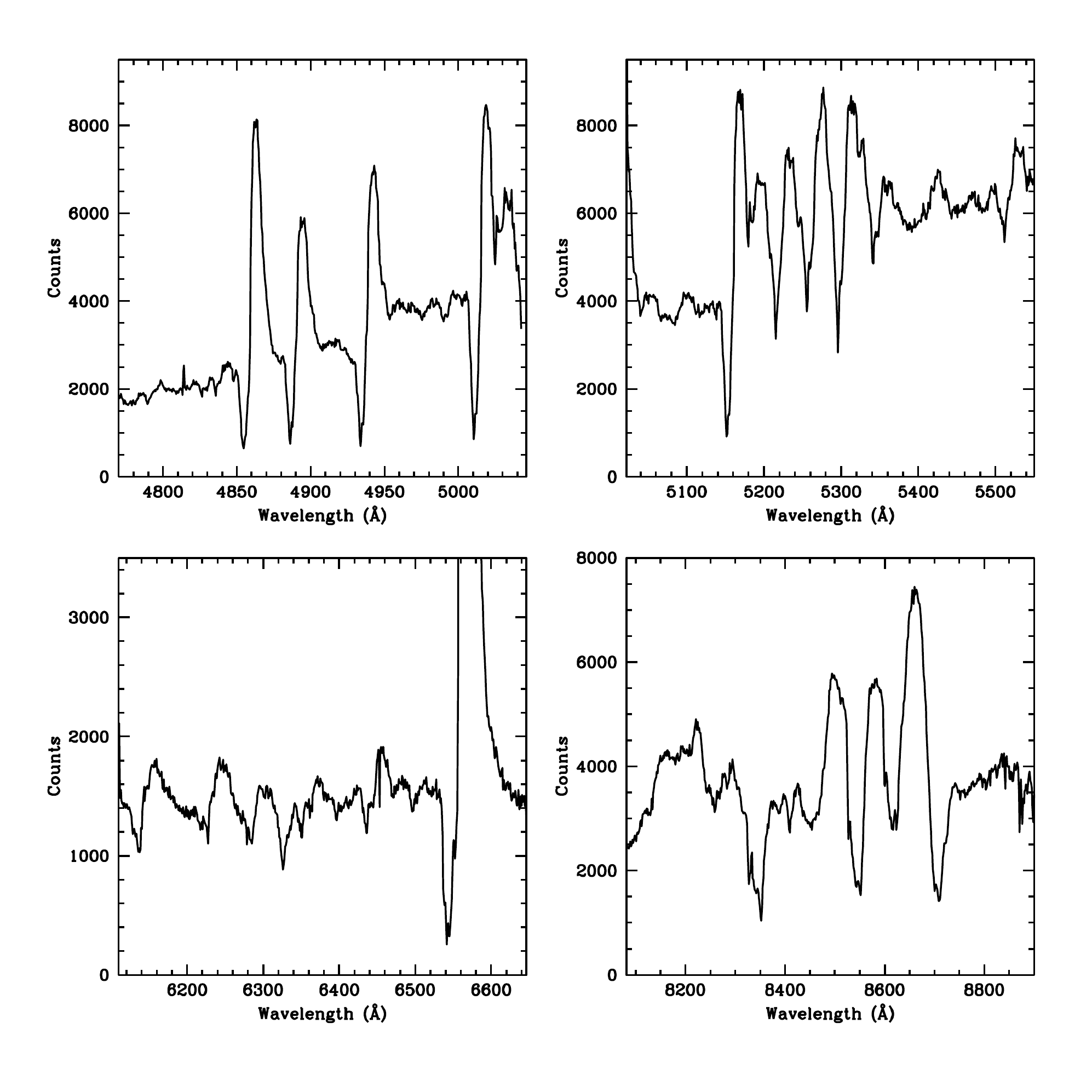}}}
\caption{The spectrum of V992 Sco on 1992 August 21.}
\label{scoaug}
\end{figure}

We use DUSTY to model the last set of our near-IR data for V992 Sco, that for 
JD 2448876.  We also have optical data (Table \ref{scophot}) that was obtained 
four days earlier. The SED for this date is plotted in Fig. \ref{jd8876}. The 
best fitting DUSTY model has a central source temperature of T$_{\rm eff}$ = 
10,000 K, a dust condensation temperature of 1050 K, a normal MRN grain 
distribution, but a geometrically thick dust shell, where the dust density 
falls off slowly as $\rho \propto$ r$^{\rm -1}$. The optical depth in the dust 
shell is $\tau$ = 2.9. Clearly, V992 Sco produced an extensive dust shell. 
Smith et al.  (1995) found that they needed to include both silicates and SiO 
emission into their models to recreate the mid-infrared spectra that they 
obtained. Here we only used amorphous carbon. The addition of silicates did not 
enhance the fit to our data set. V992 Sco was by far the brightest CNe of our 
tenure at MSSSO, and one of the brightest CNe in the IR ever. It is unfortunate 
that our time at MSSSO ended before the outburst of V992 Sco was complete.

\renewcommand{\thefigure}{22}
\begin{figure}
\centerline{{\includegraphics[width=7cm]{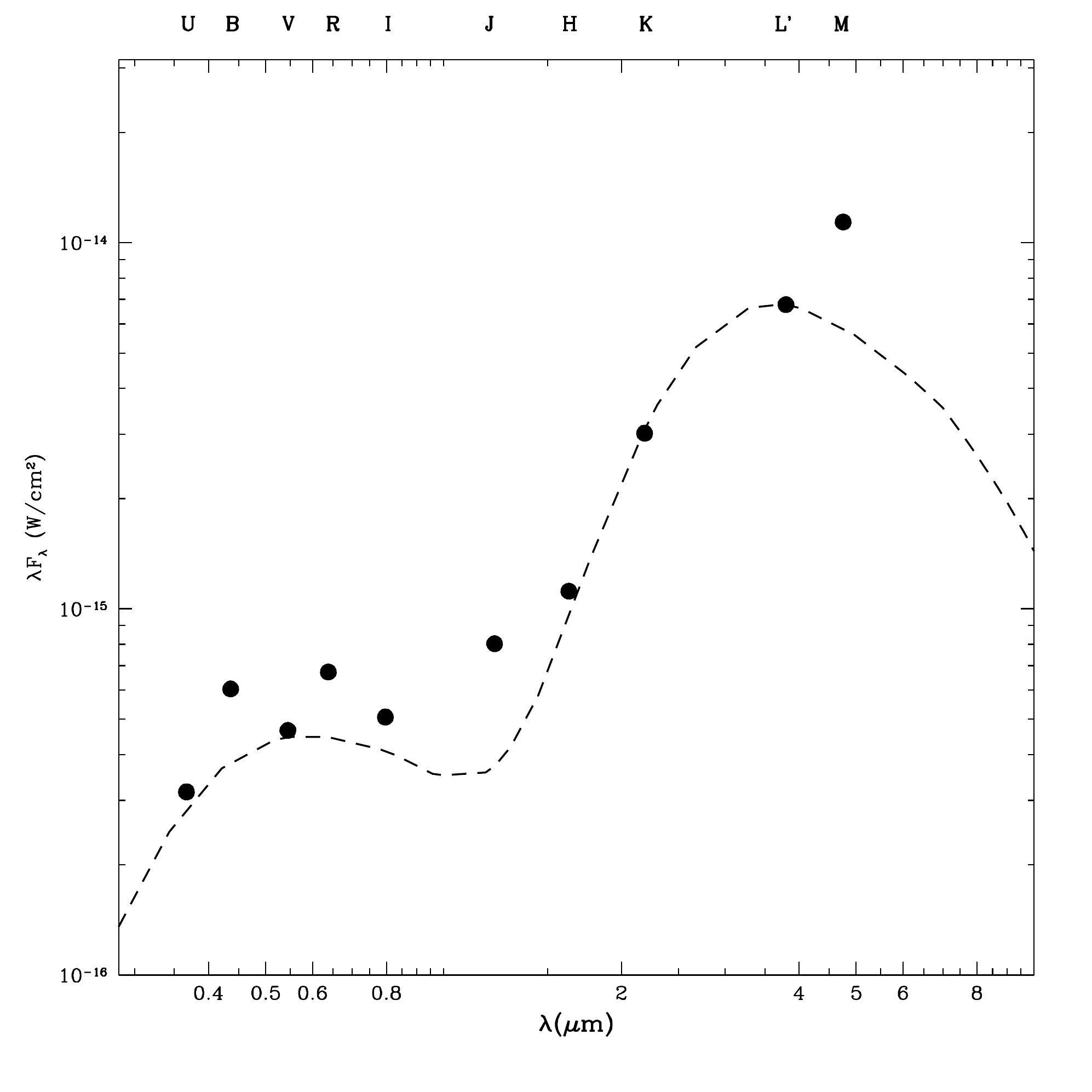}}}
\caption{The SED of V992 Sco for JD 2448876. The dashed line is the dust
shell model described in the text.}
\label{jd8876}
\end{figure}

\begin{flushleft}
{\bf V4157 Sgr:} The final CNe of our program was V4157 Sgr, discovered
by W. Liller and P. Camilleri on 1992 February 13 (IAUC 5451). We did not
begin observing it until two months later. Our two sets of observations were
obtained using a 7" aperture due to a $K$ = 9.9 mag star located very
close to this object. As demonstrated by the broad H I emission lines
in its infrared spectrum, Fig. \ref{sgr92}, V4157 Sgr was a very fast
CNe, and Williams et al. (2003) list $t_{\rm 3}$ = 12 d. The SED is 
consistent with one dominated by line emission. Smith et al. obtained
mid-IR spectrum on JD 2448759, and found a free-free emission source.
\end{flushleft}

\renewcommand{\thefigure}{23}
\begin{figure}
\centerline{{\includegraphics[width=7cm]{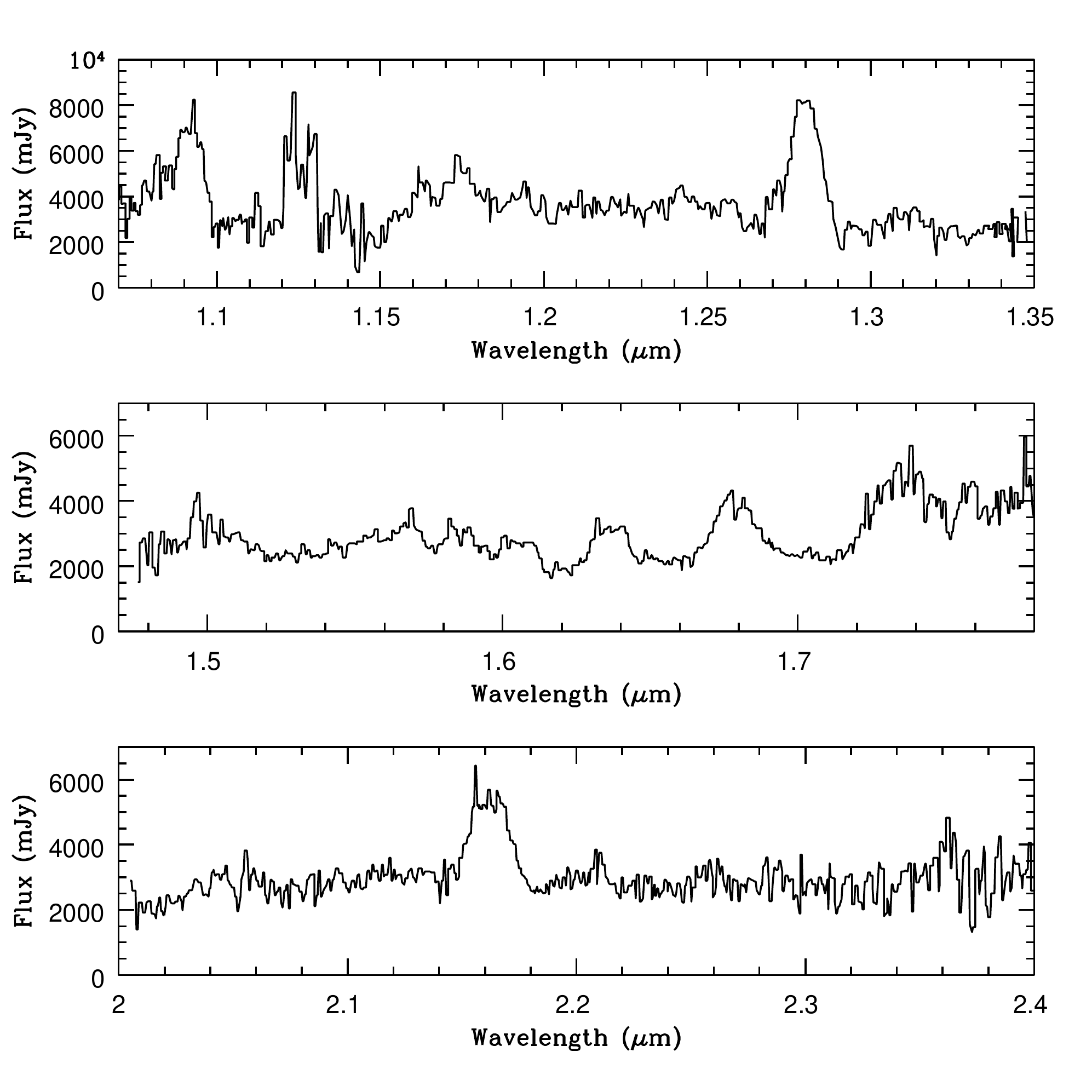}}}
\caption{The near-IR spectra of V4157 Sgr obtained using CIGS on the
SSO 2.3 m telescope on 1992 June 23.}
\label{sgr92}
\end{figure}

\section{Discussion and Conclusions}

With nearly monthly access to an infrared photometer on a 2.3 m telescope,
in the spring of 1991, we began a program of following-up recently discovered 
CNe to monitor them for the production of dust in their ejecta. We combined
this with occasional optical photometry, and optical and infrared spectroscopy.
Of the ten CNe discussed in this paper, five of them clearly produced dust. Of 
course, four of these five objects just happen to be the CNe with the most 
extensive photometric coverage, and we can only really exclude dust production 
for V351 Pup. As far as we can tell, our $M$-band spectrum of V992 Sco is the 
only existing spectrum that shows the CO fundamental in emission for a CNe. 
It is proof that the $M$-band excesses seen in many dust-producing CNe is due 
to CO emission.

Even though V992 Sco and V868 Cen produced significant amounts of dust, neither
produced shells with optical depths larger than $\tau$ $=$ 3. It is likely
that geometry plays a larger role in the obscuration of the central sources
in CNe than the quantity of dust produced in their ejecta. Dust shell masses
calculated assuming opaque spherical shells almost certainly over-estimate
the true dust masses produced by CNe.

The real question that has yet to be answered is why do some CNe produce
dust, while others do not? Our small sample says the rate of dust production
may be as high as 50\%. Perhaps the $\sim$ 18 month period during which we were 
observing was biased towards Fe II-type CNe, and these preferentially produce 
dust. It will be hard to know without a larger survey. But this will require
both optical spectroscopy for classification, and infrared photometry for
dust shell detection.
Unfortunately, as imperfect as our survey was, we do not see it being
repeated anytime soon. As shown for both V2264 Oph and Nova Oph 1991b, it is 
essential to get data beyond the $K$-band to confirm the presence of dust when 
you are unable to obtain good temporal coverage---and given the difficulty of
convincing a TAC to allocate any time to observe CNe, adequate temporal coverage 
will be hard to come by. Getting beyond the $K$-band will be even harder!

\acknowledgements We need to posthumously acknowledge the help of Peter
McGregor and Alex Rodgers. Alex allocated the money to buy the
electronics needed to drive the InSb array for the upgrade to CIGS, and Peter
wrote the software to acquire and (eventually) reduce the data from CIGS
on the fly. {\it If only the weather had been slightly more cooperative...} We 
also acknowledge that several of the nights of data for V992 
Sco were {\it likely} obtained by G. Stringfellow. Though it is so long ago, 
it is difficult to remember who observed what, when.

\section{References}
\begin{flushleft}
Bornak, J. 2012, PhD Thesis, New Mexico State University\\
Cardelli, J. A., Clayton, G. C., \& Mathis, J. S. 1989, ApJ, 345, 245\\
Della Valle, M., \& Livio, M. 1995, ApJ, 452, 704\\
Downes, R. A., \& Duerbeck, H. W. 2000, AJ, 120, 2007\\
Gordon, K. D., Misslet, K. A., Witt, A. N., Clayton, G. C. 2001, AJ, 551, 269\\
Harrison, T. E., Bornak, J., McArthur, B. E., \& Benedict, G. F. 2013, ApJ, 767, 7\\
Harrison, T. E., \& Stringfellow, G. S. 1994, 437, 827\\
Ivezi\'{c}, \^{Z}., Nenkova, M., Heymann, F. \& Elitzur, M., 2012, ``User Manual 
for DUSTY (V4)''\\
Jones, T. J., Hyland, A. R., Dopita, M. A., Hart, J., Conroy, P., Hillier, J.
1982, PASP, 94, 207\\
Mathis, J. S., Rumpl, W., \& Nordsieck, K. H. 1977, ApJ, 217, 425\\
McGregor, P. J. 1994, PASP, 106, 508\\
Osterbrock, D. E. 1989, Astrophysics of gaseous nebulae and active galactic 
nuclei, ed.  Osterbrock, D. E.\\
Shara, M. M., Doyle, T. F., Pagnotta, A., Garland, J. T., Lauer, T. R., et al.
2018, MNRAS, 474, 1746\\
Shara, M. M. 2014, ASPC, 490, 3\\
Shore, S. N. 2013, A\&A, 559, L7\\
Smith, C. H., Aitken, D. K., Roche, P. F., \& Wright, C. M. 1995, MNRAS, 277, 259\\
Starrfield  S., Iliadis, C., \& Hix, R. 2008, in Classical Novae, ed. M. Bode \&
A. Evans (2nd ed.; Cambridge: Cambridge Univ. Press), 77\\
van Genderen, A. M. 1992, A\&A, 257, 177\\
Venturini, C. C., Rudy, R. J., Lynch, D. K., Mazuk, S., \& Peutter, R. C. 2004,
AJ, 128, 405\\
Williams, R. E., Hamuy, M., Phillips, M. M., Heathcote, S. R., Wells, L., et al.
2003, JAD, 9, 3\\
Williams, R. E., Phillips, M. M., \& Hamuy, M. 1994, ApJS, 90, 297\\
Yaron, O., Prialnik, D., Shara, M. M., \& Kovetz, A. 2005, ApJ, 623, 398\\
\end{flushleft}

\begin{deluxetable}{lccccc}
\tablecolumns{6}
\tablewidth{0pt}
\centering
\tablecaption{Infrared Photometry of Classical Novae}
\tablehead{
\colhead{Julian Date} & \colhead{$J$} & \colhead{$H$} & \colhead{K} & \colhead{L'} & \colhead{M}}
\startdata
\cutinhead{Nova Cen 1991 = V868 Cen}
2448375.00& 6.28 $\pm$ 0.05& 5.69 $\pm$ 0.05&5.17 $\pm$ 0.05&4.53 $\pm$ 0.10& \nodata  \\
2448376.00& 6.18 $\pm$ 0.05& 5.52 $\pm$ 0.05&5.06 $\pm$ 0.05&4.61 $\pm$ 0.10& \nodata  \\
2448377.00& 6.16 $\pm$ 0.05& 5.53 $\pm$ 0.05&5.09 $\pm$ 0.05&4.61 $\pm$ 0.10& \nodata  \\
2448379.00& 6.21 $\pm$ 0.05& 5.46 $\pm$ 0.05&4.90 $\pm$ 0.05&4.42 $\pm$ 0.10& \nodata  \\
2448380.00& 5.97 $\pm$ 0.05& 5.26 $\pm$ 0.04&4.73 $\pm$ 0.04&4.22 $\pm$ 0.06& \nodata  \\
2448381.00& 6.20 $\pm$ 0.04& 5.47 $\pm$ 0.04&4.89 $\pm$ 0.04&4.10 $\pm$ 0.06& \nodata  \\
2448382.00& 6.22 $\pm$ 0.04& 5.56 $\pm$ 0.04&5.00 $\pm$ 0.04&4.32 $\pm$ 0.06&3.20 $\pm$ 0.10 \\
2448428.98& 6.69 $\pm$ 0.03& 5.66 $\pm$ 0.03&4.28 $\pm$ 0.03&2.48 $\pm$ 0.06&1.62 $\pm$ 0.10 \\
2448429.81& 6.65 $\pm$ 0.03& 5.69 $\pm$ 0.03&4.44 $\pm$ 0.03&2.58 $\pm$ 0.06&1.40 $\pm$ 0.10 \\
2448490.86& 9.26 $\pm$ 0.03& 6.98 $\pm$ 0.03&4.51 $\pm$ 0.03&1.92 $\pm$ 0.06&0.61 $\pm$ 0.10 \\
2448492.90& 9.20 $\pm$ 0.02& 6.84 $\pm$ 0.02&4.46 $\pm$ 0.02&1.81 $\pm$ 0.03&0.83 $\pm$ 0.10 \\
2448494.86& 9.14 $\pm$ 0.02& 6.82 $\pm$ 0.02&4.44 $\pm$ 0.02&1.91 $\pm$ 0.03&0.70 $\pm$ 0.10 \\
2448522.71& 8.73 $\pm$ 0.10& 6.83 $\pm$ 0.03&4.68 $\pm$ 0.03&2.45 $\pm$ 0.06&1.40 $\pm$ 0.10 \\
2448670.04&10.82 $\pm$ 0.02& 9.26 $\pm$ 0.02&7.28 $\pm$ 0.02&5.05 $\pm$ 0.02& \nodata  \\
2448674.04&10.87 $\pm$ 0.02& 9.37 $\pm$ 0.02&7.38 $\pm$ 0.02&5.13 $\pm$ 0.03& \nodata  \\
2448741.01&11.59 $\pm$ 0.04&11.24 $\pm$ 0.02&8.35 $\pm$ 0.03&6.07 $\pm$ 0.04&5.07 $\pm$ 0.29 \\
2448793.05&12.13 $\pm$ 0.05&11.05 $\pm$ 0.05&9.30 $\pm$ 0.03&6.95 $\pm$ 0.05&5.75 $\pm$ 0.30 \\
2450963.58&16.81 $\pm$ 0.15&16.10 $\pm$ 0.12&15.48 $\pm$ 0.11& \nodata & \nodata \\
\cutinhead{Nova Oph 1991 a = V2264 Oph}
2448375.10& 9.19 $\pm$ 0.05& 8.92 $\pm$ 0.05& 8.52 $\pm$ 0.05&7.84 $\pm$ 0.10& \nodata \\
2448376.10& 9.20 $\pm$ 0.05& 8.93 $\pm$ 0.05& 8.50 $\pm$ 0.05&7.79 $\pm$ 0.10& \nodata  \\
2448377.10& 9.19 $\pm$ 0.05& 8.96 $\pm$ 0.05& 8.52 $\pm$ 0.05&7.81 $\pm$ 0.10& \nodata  \\
2448379.10& 9.19 $\pm$ 0.05& 9.03 $\pm$ 0.05& 8.60 $\pm$ 0.05&7.87 $\pm$ 0.10& \nodata  \\
2448380.10& 9.51 $\pm$ 0.05& 9.57 $\pm$ 0.05& 8.77 $\pm$ 0.04&8.05 $\pm$ 0.07& \nodata  \\
2448381.10& 9.32 $\pm$ 0.04& 9.13 $\pm$ 0.04& 8.75 $\pm$ 0.04&8.02 $\pm$ 0.07& \nodata  \\
2448382.10& 9.36 $\pm$ 0.04& 9.20 $\pm$ 0.04& 8.77 $\pm$ 0.04&8.11 $\pm$ 0.06& \nodata  \\
2448397.99&10.42 $\pm$ 0.03&10.57 $\pm$ 0.03& 9.72 $\pm$ 0.03&7.81 $\pm$ 0.06& \nodata  \\
2448430.12&10.54 $\pm$ 0.03&10.70 $\pm$ 0.03& 9.73 $\pm$ 0.03&9.08 $\pm$ 0.06& \nodata  \\
2448490.92&11.31 $\pm$ 0.03&11.52 $\pm$ 0.03&10.69 $\pm$ 0.03&9.72 $\pm$ 0.06& \nodata  \\
2448492.95&11.45 $\pm$ 0.03&11.74 $\pm$ 0.03&10.77 $\pm$ 0.03&9.87 $\pm$ 0.17& \nodata  \\
2448494.95&11.39 $\pm$ 0.02&11.64 $\pm$ 0.02&10.79 $\pm$ 0.02&10.09 $\pm$ 0.20& \nodata  \\
2448522.92&11.76 $\pm$ 0.04&11.85 $\pm$ 0.04&10.95 $\pm$ 0.04&10.26 $\pm$ 0.15& \nodata  \\
2448524.00&11.76 $\pm$ 0.03&11.88 $\pm$ 0.03&11.03 $\pm$ 0.03&9.60 $\pm$ 0.10& \nodata  \\
2448526.01&11.66 $\pm$ 0.03&11.86 $\pm$ 0.03&10.98 $\pm$ 0.03&10.13 $\pm$ 0.22& \nodata  \\
2448732.09&13.49 $\pm$ 0.07&13.00 $\pm$ 0.06&12.76 $\pm$ 0.13& 9.22 $\pm$ 0.14& \nodata  \\
2448756.00&14.31 $\pm$ 0.06&13.94 $\pm$ 0.05&12.86 $\pm$ 0.05& 9.37 $\pm$ 0.25&$>$5.5  \\
2448796.04&14.19 $\pm$ 0.04&14.32 $\pm$ 0.06&13.20 $\pm$ 0.07& 9.65 $\pm$ 0.18& \nodata  \\
2448796.15&14.14 $\pm$ 0.08&13.87 $\pm$ 0.10&12.91 $\pm$ 0.05& 9.73 $\pm$ 0.15& \nodata  \\
2450963.80&18.40 $\pm$ 0.30& $>$ 18 & 17.02 $\pm$  0.28& \nodata & \nodata \\
\cutinhead{Nova Oph 1991 b}
2448732.20&14.82 $\pm$ 0.28&14.64 $\pm$ 0.26&11.10 $\pm$ 0.04& 8.08 $\pm$ 0.05& \nodata \\
2448756.05&\nodata &14.35 $\pm$ 0.27 & 11.48 $\pm$ 0.04 & 8.76 $\pm$ 0.11& \nodata \\
2448796.16&15.18 $\pm$ 0.19&14.92 $\pm$ 0.19&12.20 $\pm$ 0.05&9.59 $\pm$ 0.27& \nodata\\
\cutinhead{Nova Pup 1991 = V351 Pup}
2448670.04& 7.62 $\pm$ 0.02& 7.78 $\pm$ 0.02& 7.03 $\pm$ 0.02& 6.42 $\pm$ 0.03& \nodata \\
2448675.02& 7.70 $\pm$ 0.02& 7.91 $\pm$ 0.04& 7.16 $\pm$ 0.02& 6.55 $\pm$ 0.02& \nodata \\
2448731.91& 8.30 $\pm$ 0.04& 8.62 $\pm$ 0.02& 7.92 $\pm$ 0.03& 7.12 $\pm$ 0.04& \nodata\\
2448733.90& 8.36 $\pm$ 0.03& 8.65 $\pm$ 0.02& 7.94 $\pm$ 0.03& 7.19 $\pm$ 0.07& \nodata\\
2448756.20& 8.76 $\pm$ 0.03& 8.95 $\pm$ 0.03& 8.25 $\pm$ 0.03& 7.46 $\pm$ 0.04& \nodata\\
2448792.88& 9.06 $\pm$ 0.05& 9.26 $\pm$ 0.05& 8.50 $\pm$ 0.03& 7.57 $\pm$ 0.08& \nodata\\
2448794.86& 8.98 $\pm$ 0.04& 9.22 $\pm$ 0.04& 8.47 $\pm$ 0.04& 7.57 $\pm$ 0.04& \nodata\\
2450963.60&19.85 $\pm$ 0.28&18.25 $\pm$ 0.33& $>$ 17.5 & \nodata & \nodata \\
\cutinhead{Nova Sgr 1991 = V4160 Sgr}
2448490.99&11.50 $\pm$ 0.03&12.04 $\pm$ 0.03&10.86 $\pm$ 0.03& 9.53 $\pm$ 0.18& \nodata\\
2448493.02&11.92 $\pm$ 0.02&12.28 $\pm$ 0.02&10.95 $\pm$ 0.02& 9.87 $\pm$ 0.17& \nodata\\
2448495.00&11.92 $\pm$ 0.03&12.39 $\pm$ 0.04&11.05 $\pm$ 0.03&10.09 $\pm$ 0.20& \nodata\\
\cutinhead{Nova Sct 1991 = V444 Sct}
2448523.98&10.68 $\pm$ 0.03&10.48 $\pm$ 0.03&9.89 $\pm$ 0.03&9.23 $\pm$ 0.08& \nodata\\
2448526.04&10.87 $\pm$ 0.03&10.72 $\pm$ 0.03&10.15 $\pm$ 0.03& 9.51 $\pm$ 0.10& \nodata\\
2448756.25&15.46 $\pm$ 0.22&14.48 $\pm$ 0.12&13.99 $\pm$ 0.05&10.92 $\pm$ 0.32& \nodata\\
2448796.24&16.17 $\pm$ 0.29& \nodata & 14.77 $\pm$ 0.22 & \nodata & \nodata  \\
\cutinhead{Nova LMC 1991 a}
2448380.80&10.69 $\pm$ 0.04&10.82 $\pm$ 0.04&10.34 $\pm$ 0.04&9.36 $\pm$ 0.09& \nodata\\
2448381.80&10.82 $\pm$ 0.04&10.74 $\pm$ 0.04& 9.83 $\pm$ 0.04&8.76 $\pm$ 0.06& \nodata\\
~ \\
\cutinhead{Nova Sco 1992 = V992 Sco}
2448793.07&6.05 $\pm$ 0.04&5.70 $\pm$ 0.04&5.32 $\pm$ 0.03&4.59 $\pm$ 0.04&3.40 $\pm$ 0.10\\
2448794.07&6.01 $\pm$ 0.03&5.73 $\pm$ 0.03&5.30 $\pm$ 0.03&4.61 $\pm$ 0.03&3.39 $\pm$ 0.10\\
2448795.08&5.90 $\pm$ 0.03&5.60 $\pm$ 0.03&5.27 $\pm$ 0.04&4.46 $\pm$ 0.03&3.36 $\pm$ 0.10\\
2448796.13&5.93 $\pm$ 0.02&5.58 $\pm$ 0.02&5.09 $\pm$ 0.02&4.44 $\pm$ 0.02&3.26 $\pm$ 0.10\\
2448797.08&5.55 $\pm$ 0.03&5.21 $\pm$ 0.03&4.85 $\pm$ 0.03&4.35 $\pm$ 0.03&3.23 $\pm$ 0.10\\
2448814.08&5.80 $\pm$ 0.10&5.66 $\pm$ 0.10&5.32 $\pm$ 0.10&4.33 $\pm$ 0.10&\nodata \\
2448816.08&5.71 $\pm$ 0.10&5.30 $\pm$ 0.10&4.95 $\pm$ 0.10&4.38 $\pm$ 0.10&\nodata \\
2448820.08&5.27 $\pm$ 0.10&4.80 $\pm$ 0.10&4.38 $\pm$ 0.10&3.83 $\pm$ 0.10&\nodata \\
2448846.08&5.25 $\pm$ 0.02&4.55 $\pm$ 0.02&3.50 $\pm$ 0.02&1.78 $\pm$ 0.04&0.98 $\pm$ 0.10\\
2448876.08&7.10 $\pm$ 0.04&5.81 $\pm$ 0.02&3.80 $\pm$ 0.02&1.16 $\pm$ 0.05&$-$0.13 $\pm$ 0.10\\
2450963.68&11.93 $\pm$ 0.03&10.36 $\pm$ 0.03& 9.81 $\pm$ 0.03& \nodata & \nodata\\
\cutinhead{Nova Sgr 1992 = V4157 Sgr}
2448732.20&11.23 $\pm$ 0.03&11.27 $\pm$ 0.02&10.25 $\pm$ 0.03&8.79 $\pm$ 0.20&\nodata\\
2448796.19&12.04 $\pm$ 0.03&12.07 $\pm$ 0.03&11.15 $\pm$ 0.03&9.96 $\pm$ 0.18&\nodata\\
\enddata
\label{irphot}
\end{deluxetable}
\clearpage
\begin{deluxetable}{lccccc}
\tablecolumns{6}
\tablewidth{0pt}
\centering
\tablecaption{Optical Photometry of Classical Novae}
\tablehead{\colhead{Julian Date} & \colhead{($U - B$)}&\colhead{($B - V$)}& \colhead{$V$}&\colhead{($V - R$)}&\colhead{($V - I$)}}
\startdata
\cutinhead{Nova Her 1991 = V838 Her}
2448349.70&$-$0.84 $\pm$ 0.10&$-$0.01 $\pm$ 0.03&10.58 $\pm$ 0.02&1.66 $\pm$ 0.03&1.24 $\pm$ 0.03\\
2448352.68&$-$0.77 $\pm$ 0.10&$-$0.11 $\pm$ 0.03&11.45 $\pm$ 0.02&1.63 $\pm$ 0.03&1.71 $\pm$ 0.03\\
2448353.71&$-$0.86 $\pm$ 0.10&$-$0.19 $\pm$ 0.03&11.73 $\pm$ 0.02&1.61 $\pm$ 0.03&1.78 $\pm$ 0.03\\
2448450.51&$-$1.51 $\pm$ 0.20&$-$0.17 $\pm$ 0.15&15.87 $\pm$ 0.05&0.52 $\pm$ 0.05&0.71 $\pm$ 0.05\\
2448737.27&\nodata           &   0.71 $\pm$ 0.25&18.26 $\pm$ 0.15&0.86 $\pm$ 0.18&1.35 $\pm$ 0.18\\
\cutinhead{Nova Oph 1991a = V2264 Oph}
2448737.14&$-$1.091 $\pm$ 0.20&  0.57 $\pm$ 0.15&15.92 $\pm$ 0.05&0.29 $\pm$ 0.05&0.61 $\pm$ 0.05\\
2450841.60&\nodata&\nodata&21.24 $\pm$ 0.04& 1.27 $\pm$ 0.04 & 2.49 $\pm$ 0.06\\
\cutinhead{Nova Oph 1991b}
2448737.19&15.69 $\pm$ 0.05&$-$0.031 $\pm$ 0.20&  1.40 $\pm$ 0.15&$-$0.21$\pm$ 0.05&$-$0.39 $\pm$ 0.05\\
\cutinhead{Nova Sgr 1992 = V4157 Sgr}
2448737.25&13.47 $\pm$ 0.03&$-$0.22  $\pm$ 0.15&  0.71 $\pm$ 0.10&1.38 $\pm$ 0.04&1.49 $\pm$ 0.04\\

\enddata
\label{cneopt}
\end{deluxetable}
\clearpage
\begin{deluxetable}{lccccccc}
\tablecolumns{8}
\tablewidth{0pt}
\centering
\tablecaption{Optical Photometry of V868 Cen}
\tablehead{\colhead{Julian Date} & \colhead{t $-$ t$_{\rm max}$} & \colhead{($U - B$)}&\colhead{($B - V$)}& \colhead{$V$}&\colhead{($V - R$)}&\colhead{($V - I$)} & \colhead{Source}}
\startdata
2448350.50&  4.50&\nodata &\nodata & 10.34&\nodata &\nodata  &1$^{\rm a}$\\
2448350.65&  4.65&\nodata &\nodata & 10.54&\nodata &\nodata  &1$^{\rm a}$\\
2448352.04&  6.04& 0.33   & 1.66   & 10.69& 1.57   & 2.20    &1$^{\rm b}$\\
2448352.61&  6.61&\nodata &\nodata & 10.64&\nodata &\nodata  &1$^{\rm a}$\\
2448352.68&  6.68&\nodata &\nodata & 10.74&\nodata &\nodata  &1$^{\rm a}$\\
2448354.02&  8.02& 0.31   & 1.57   & 10.89& 1.66   & 2.18    &1$^{\rm b}$\\
2448356.61& 10.61&\nodata &\nodata & 10.84&\nodata &\nodata  &1$^{\rm a}$\\
2448359.88& 13.88&\nodata & 1.60   & 12.00& 1.50   & 3.00    &2$^{\rm b}$\\
2448361.62& 15.62&\nodata &\nodata & 10.54&\nodata &\nodata  &1$^{\rm a}$\\
2448363.52& 17.52&\nodata &\nodata & 10.44&\nodata &\nodata  &1$^{\rm a}$\\
2448364.53& 18.53&\nodata &\nodata & 10.34&\nodata &\nodata  &1$^{\rm a}$\\
2448365.52& 19.52&\nodata &\nodata & 10.14&\nodata &\nodata  &1$^{\rm a}$\\
2448366.50& 20.50&\nodata &\nodata & 10.24&\nodata &\nodata  &1$^{\rm a}$\\
2448367.60& 21.60&\nodata &\nodata & 10.24&\nodata &\nodata  &1$^{\rm a}$\\
2448368.13& 22.13& 0.33   & 1.58   & 10.97& 1.3    & 2.42    &3$^{\rm b}$\\
2448368.53& 22.53&\nodata &\nodata & 10.24&\nodata &\nodata  &1$^{\rm a}$\\
2448370.51& 24.51&\nodata &\nodata & 10.14&\nodata &\nodata  &1$^{\rm a}$\\
2448371.48& 25.48&\nodata &\nodata & 10.04&\nodata &\nodata  &1$^{\rm a}$\\
2448375.49& 29.49&\nodata &\nodata &  9.84&\nodata &\nodata  &1$^{\rm a}$\\
2448377.03& 31.03& 0.56   & 1.67   & 10.27& 1.17   & 2.27    &4$^{\rm b}$\\
2448377.50& 31.50&\nodata &\nodata &  9.74&\nodata &\nodata  &1$^{\rm a}$\\
2448378.10& 32.10& 0.44   & 1.67   & 10.24& 1.21   & 2.41    &4$^{\rm b}$\\
2448379.68& 33.68&\nodata &\nodata &  9.84& 1.40   &\nodata  &1$^{\rm a}$\\
2448381.52& 35.52&\nodata &\nodata & 10.04&\nodata &\nodata  &1$^{\rm a}$\\
2448383.53& 37.53&\nodata &\nodata & 10.04&\nodata &\nodata&1$^{\rm a}$\\
2448388.67& 42.67&\nodata &\nodata & 10.74&\nodata &\nodata&1$^{\rm a}$\\
2448389.50& 43.50&\nodata &\nodata & 10.84&\nodata &\nodata&1$^{\rm a}$\\
2448413.49& 67.49&\nodata &\nodata & 11.04&\nodata &\nodata&1$^{\rm a}$\\
2448451.93& 105.9&\nodata & 1.74   & 13.74&\nodata &\nodata&5$^{\rm a}$\\
2448472.00& 126.0&\nodata & 1.60   & 13.38&\nodata &\nodata&6$^{\rm a}$\\
2448498.91& 152.9&\nodata & 1.30   & 14.25&\nodata &\nodata&7$^{\rm a}$\\
2448502.87& 156.9&\nodata & 1.04   & 14.09& 1.96   & 2.64& 1$^{\rm b}$\\
2448608.10& 262.1&\nodata & 1.06   & 14.72&\nodata & \nodata&8$^{\rm a}$\\
2448666.22& 320.2& 1.40   & 1.46   & 13.55&0.94 & 1.43&1$^{\rm b}$\\
2448668.23& 322.2& 1.07   & 1.45   & 13.56&0.93 & 1.43&1$^{\rm b}$\\
2448671.06& 325.1& 0.72   & 1.61   & 13.63&0.97 & 1.45&1$^{\rm b}$\\
2448737.10& 391.1& 1.34   & 1.52   & 13.90&0.65 & 1.15&1$^{\rm b}$\\
2450963.58&2617.6& 0.90   & 1.43   & 18.85&0.9 & 1.59&1$^{\rm b}$\\
\enddata
\tablenotetext{a}{Errors on photometry are $\pm$ 0.1 mag},\tablenotetext{b}{Errors on $BVRI$
photometry are $\pm$ 0.05 mag, errors in $U$ are $\pm$ 0.21 mag.}
\tablerefs{1) Bornak (2012), 2) Cappellaro et al. (1991), 3) McNaught et al. 
(1991b), 4) Jablonski et al.  (1991a), 5) Gilmore (1991b), 6) Gilmore (1991a), 
7) Gilmore (1991c), 8) Kilmarten (1992). Note that much of the visual (only)
data in Bornak (2012) was provided by W. Liller.}
\label{cenphot}
\end{deluxetable}

\begin{deluxetable}{lccccc}
\tablecolumns{6}
\tablewidth{0pt}
\centering
\tablecaption{Optical Data for V992 Sco}
\tablehead{\colhead{Julian Date} & \colhead{($U - B$)}&\colhead{($B - V$)}& \colhead{$V$}&\colhead{($V - R$)}&\colhead{($V - I$)}}
\startdata
2448741.10&\nodata&\nodata&$>$ 12&\nodata&\nodata\\
2448753.00&\nodata&\nodata&{\it 10} &\nodata&\nodata\\
2448764.99&\nodata&\nodata&{\it 8.2} &\nodata&\nodata\\
2448766.01&\nodata&\nodata&{\it 8.3}&\nodata&\nodata\\
2448766.87&\nodata&\nodata& 8.3 &\nodata&\nodata\\
2448766.92&\nodata&\nodata&{\it 8.4}&\nodata&\nodata\\
2448768.90& 0.79  & 1.19  & 7.26& 0.72  & 2.24\\
2448773.20& 0.22  & 1.15  & 8.97& 0.93  & 3.00\\
2448774.11& 0.12  & 1.08  & 9.03& 0.94  & 2.96\\
2448774.53&\nodata&\nodata& 8.6 &\nodata&\nodata\\
2448774.53&\nodata&\nodata& 8.5 &\nodata&\nodata\\
2448775.69&\nodata&\nodata& 9.0 &\nodata&\nodata\\
2448776.82&\nodata&\nodata& 9.0 &\nodata&\nodata\\
2448775.01& 0.01  & 1.07  & 9.13&\nodata&\nodata\\
2448775.69&\nodata&\nodata& 8.8 &\nodata&\nodata\\
2448775.94& 0.03  & 1.01  & 9.22&\nodata&\nodata\\
2448776.70&\nodata&\nodata& 8.22&\nodata&\nodata\\
2448776.95& 0.03  & 1.01  & 9.03&\nodata&\nodata\\
2448778.94& 0.02  & 0.95  & 9.36&\nodata&\nodata\\
2448779.47&\nodata&\nodata& 8.50&\nodata&\nodata\\
2448780.67&\nodata&\nodata& 8.60&\nodata&\nodata\\
2448782.20& 0.09  & 1.05  & 8.36&\nodata&\nodata\\
2448782.55&\nodata&\nodata& 7.66&\nodata&\nodata\\
2448783.52&\nodata&\nodata& 7.87&\nodata&\nodata\\
2448783.94& 0.07  & 1.10  & 7.93&\nodata&\nodata\\
2448783.99& 0.03  & 1.05  & 7.86& 0.76  & 1.57\\
2448784.56&\nodata&\nodata& 7.80&\nodata&\nodata\\
2448784.64&\nodata&\nodata& 8.10&\nodata&\nodata\\
2448784.89&\nodata&\nodata& 8.40&\nodata&\nodata\\
2448785.53&\nodata&\nodata& 7.90&\nodata&\nodata\\
2448785.69&\nodata&\nodata& 7.41&\nodata&\nodata\\
2448785.92& 0.04  & 1.06  & 8.15&\nodata&\nodata\\
2448786.47&\nodata&\nodata& 8.20&\nodata&\nodata\\
2448787.87& $-$0.06 & 0.94& 8.24& 0.83   & 1.53\\
2448789.95& $-$0.19 & 0.76& 8.79&\nodata&\nodata\\
2448790.02&\nodata&\nodata& 8.70&\nodata&\nodata\\
2448793.92&\nodata&\nodata& 8.90&\nodata&\nodata\\
2448794.91&\nodata&\nodata& 8.90&\nodata&\nodata\\
2448797.85& $-$0.11& 1.02 & 8.16&\nodata&\nodata\\
2448798.07& $-$0.07& 1.02 & 8.05& 0.79  & 1.51\\
2448798.87& $-$0.03& 1.04 & 8.05&\nodata&\nodata\\
2448799.87& $-$0.05& 1.06 & 8.03&\nodata&\nodata\\
2448803.87& $-$0.05& 1.00 & 8.03&\nodata&\nodata\\
2448804.80& $-$0.03& 1.01 & 7.77&\nodata&\nodata\\
2448805.09& 0.01 & 1.01  & 7.62& 0.76  & 1.50\\
2448809.83& $-$0.14& 0.82& 9.34&\nodata&\nodata\\
2448809.90& $-$0.17& 0.83& 9.27& 1.31  & 1.39\\
2448819.83& $-$0.16& 1.00& 8.26&\nodata&\nodata\\
2448820.95& $-$0.13& 0.97& 8.57&\nodata&\nodata\\
2448822.00& $-$0.18& 0.92& 8.88& 1.01  & 1.66\\
2448824.05& $-$0.16& 0.90& 9.26& 1.10  & 1.57\\
2448826.85& $-$0.28& 1.03& 9.47&\nodata&\nodata\\
2448829.87& $-$0.22& 0.99& 9.46&\nodata&\nodata\\
2448830.84& $-$0.25& 1.01& 9.31&\nodata&\nodata\\
2448831.96& $-$0.21& 0.91& 9.39&\nodata&\nodata\\
2448833.93& 0.02  & 1.07  & 8.64& 0.81  & 1.54\\
2448835.88& 0.11  & 1.11  & 8.42&\nodata&\nodata\\
2448837.83& 0.18  & 1.08  & 7.91&\nodata&\nodata\\
2448840.00& 0.21  & 1.11  & 8.33& 0.78  & 1.60\\
2448867.51&\nodata&\nodata& 8.76&\nodata&\nodata\\
2448868.47&\nodata&\nodata& 8.99&\nodata&\nodata\\
2448870.50&\nodata&\nodata& 9.12&\nodata&\nodata\\
2448871.51&\nodata&\nodata& 9.33&\nodata&\nodata\\
2448872.50&\nodata&\nodata& 9.44&\nodata&\nodata\\
2448882.89& 0.30  & 0.53  &10.47& 0.96  & 1.49\\
2450963.78&\nodata&\nodata&21.03&19.97&18.75\\
\enddata
\tablerefs{Data obtained by us, or compiled from the IAU Circulars. The
error bars on most of the photometry reported in the IAU Circulars is
of order 10\%, or better. The errors on the CTIO photometry, last line, 
are $\pm$ 4\%. The italicized entries in this table are photo-visual
magnitudes.}
\label{scophot}
\end{deluxetable}
\end{document}